\documentclass[12pt, draftclsnofoot, onecolumn]{IEEEtran}
 \usepackage{amsmath,amssymb}
 \usepackage{subfigure}
 \usepackage{graphicx,graphics,color,psfrag}
 \usepackage{cite,balance}
 \usepackage{caption}
 \allowdisplaybreaks
 \usepackage{algorithm}
 \usepackage{accents}
 \usepackage{amsthm}
 \usepackage{bm}
 \usepackage{algorithmic}
 \usepackage[english]{babel}
 \usepackage{multirow}
 \usepackage{enumerate}
 \usepackage{cases}
 \usepackage{stfloats}
 \usepackage{dsfont}
 \usepackage{color,soul}
 \usepackage{amsfonts}
 \usepackage{cite,graphicx,amsmath,amssymb}
 \usepackage{subfigure}
 \usepackage{fancyhdr}
 \usepackage{hhline}
 \usepackage{graphicx,graphics}
 \usepackage{array,color}
 \usepackage{amsmath}

\newtheorem{definition}{Definition}

\newtheorem{lemma}{Lemma}
\newtheorem{corollary}{Corollary}

\newtheorem{proposition}{Proposition}

\newtheorem{remark}{\bf Remark}
\def\phi{\varphi}

\def\l{\left}
\def\r{\right}
\def\({\left(}
\def\){\right)}

\def\b0{{\mathbf{0}}}

\begin{document}
\title{\LARGE Asynchronous Mobile-Edge Computation Offloading: Energy-Efficient Resource Management} 
\author{Changsheng You,  Yong Zeng, Rui Zhang, and Kaibin Huang
  \thanks{\noindent C. You and K. Huang are  with the Department of Electrical and Electronic Engineering, The  University of  Hong Kong, Hong Kong.
    Y. Zeng and R. Zhang are with the Department of Electrical and Computer Engineering, National University of Singapore, Singapore 117583.
Corresponding author: K. Huang (Email: huangkb@eee.hku.hk). Part of this work has been accepted to IEEE ICC Workshop 2018.}
  }
\maketitle
\vspace{-40pt}
\begin{abstract}
\emph{Mobile-edge computation offloading} (MECO) is an emerging technology for enhancing mobiles' computation capabilities and prolonging their battery lifetime, by offloading intensive computation from mobiles to nearby servers such as base stations.  In this paper, we study  the energy-efficient resource-management policy for the \emph{asynchronous} MECO system, where the mobiles have \emph{heterogeneous} input-data arrival time instants and computation deadlines. First, we consider the general  case with  arbitrary arrival-deadline orders. Based on the monomial  energy-consumption model for data transmission, an optimization problem is formulated to minimize the total mobile-energy consumption under the time-sharing and computation-deadline constraints. The optimal resource-management policy for data partitioning (for offloading and local computing) and time division (for transmissions) is obtained in (semi-) closed-form expression by using the block coordinate decent method.  To gain further insights, we study the optimal resource-management design for two special cases. First, consider the case of \emph{identical} arrival-deadline orders, i.e., a mobile with input data arriving earlier also needs to complete computation earlier. The optimization problem is reduced to two sequential problems corresponding to the optimal scheduling order and  joint data-partitioning and time-division given the optimal order. It is found that the optimal time-division policy tends to equalize the defined \emph{effective computing power} among offloading mobiles via time sharing. Furthermore, this solution approach is extended to the case of \emph{reverse} arrival-deadline orders. The corresponding time-division policy is derived by a proposed \emph{transformation-and-scheduling} approach, which first determines the total offloading duration and data size for each mobile in the transformation phase and then  specifies the offloading intervals for each mobile  in the scheduling phase.
\end{abstract}
\vspace{-10pt}
\section{Introduction}
\begin{figure}[t]
\begin{center}
\includegraphics[height=5.7cm]{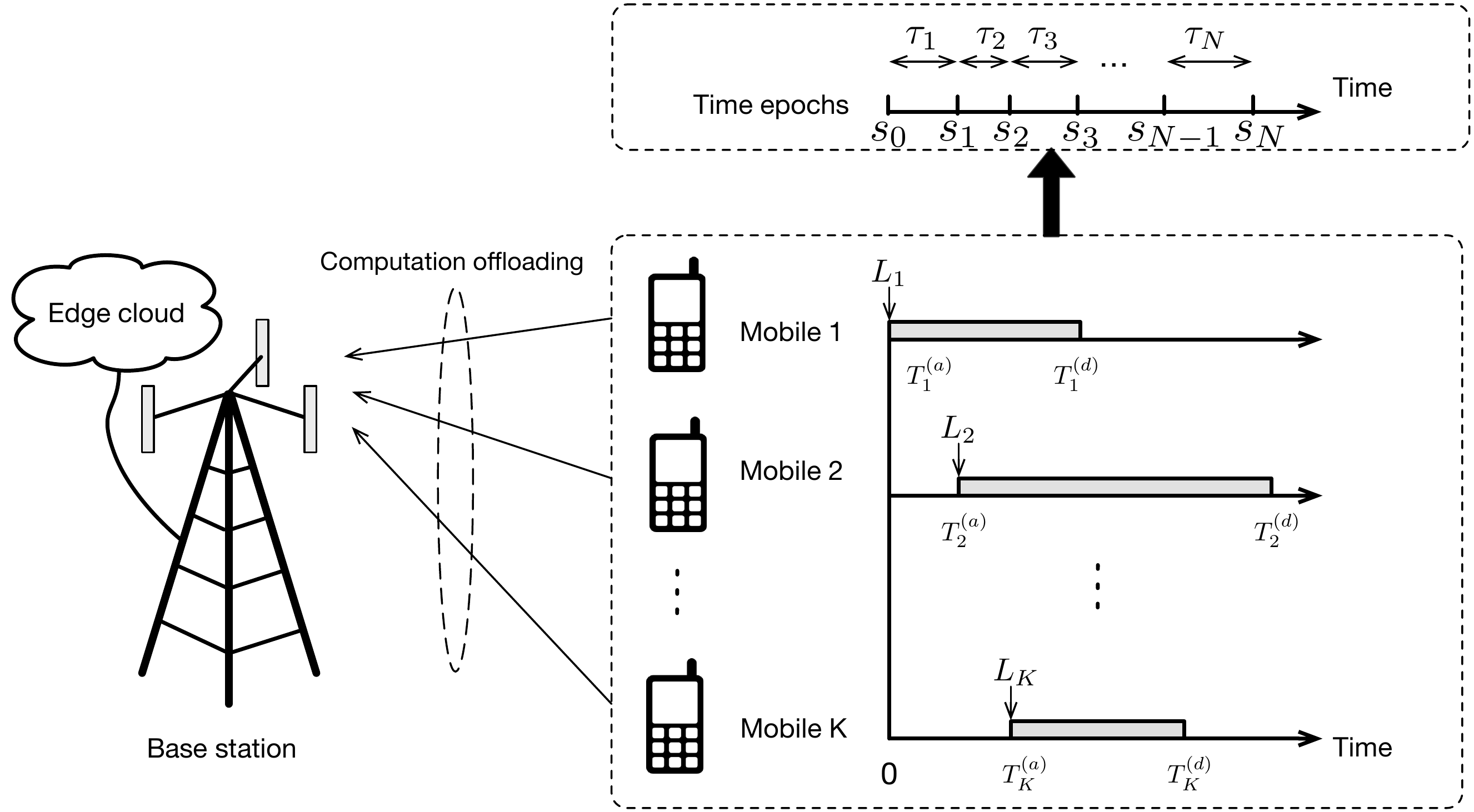}
\caption{Multiuser asynchronous  MECO systems.}
\label{Fig:System}
\end{center}
\vspace{-20pt}
\end{figure}

Realizing the vision of \emph{Internet of Things} (IoT)  has driven the unprecedented growth of small mobile devices 
 in recent years.  This stimulates  the explosive data/computation traffic  increase that is \emph{constantly}  generated from a wide range of new applications such as online gaming and video streaming. Such mobiles, however, typically suffer from finite computation capabilities and batteries due to their small form factors and low cost. Tackling these challenges gives rise to an emerging technology, called \emph{mobile-edge computation offloading} (MECO), which allows computation data to be offloaded from mobiles to proximate servers such as \emph{base stations} (BSs) and  \emph{access points} (APs), for achieving desirable low latency and mobile energy savings \cite{mao2017mobile2,taleb2017multi,wang2017survey}. In a typical  \emph{asynchrobous} MECO system as shown in Fig.~\ref{Fig:System}, 
   different mobiles generate different amounts of computation data at \emph{random} time instants, and moreover, have \emph{diverse} latency requirements depending on the applications. This complicates  the multiuser offloading and resource management in MECO systems, which shall be investigated in this work.
\vspace{-18pt}
 \subsection{Prior Work}\label{Sec:Prior}

Designing efficient MECO systems has attracted extensive attention in recent years.
In the pioneering work considering single-user MECO systems \cite{zhang:MobileMmodel:2013}, the mobile  CPU-cycle frequencies and offloading rates were optimized for maximizing the energy savings of local computing and offloading, leading to the optimal \emph{binary} offloading decision. This work was extended in \cite{you2015energyJSAC} by powering MECO with wireless energy. In addition, for applications with partitionable data, the performance of energy savings can be further enhanced by partitioning data for local computing and offloading, called \emph{partial offloading}. A set of partitioning schemes have been  proposed,  including live prefetching \cite{ko2017live}, program partitioning \cite{MahmoodiTCC16}, and controlling offloading ratio \cite{YWangTCOM16}.

The offloading design in multiuser MECO systems is more complicated. Particularly, one of the main issues is how to jointly allocate radio-and-computational resources. Most prior work on this topic assumes \emph{synchronous} MECO, where all the mobiles have the \emph{identical} data-arrival time instants and deadlines. Under this assumption,  the  resource allocation for minimizing the total mobile-energy consumption was studied in \cite{you2016energy} for both \emph{time-division multiple access} (TDMA) and \emph{orthogonal frequency-division multiple access} (OFDMA) MECO systems, where the derived optimal policy is shown to have a simple threshold-based structure.
This framework was extended in \cite{guo2016energy} to design energy-efficient multiuser MECO accounting for the non-negligible edge-cloud computing latency by using \emph{flow-shop} scheduling techniques. Further research in this direction considers more complex systems such as multi-cell MECO \cite{Barbarossa:MobileCloud:2014,nguyen2017joint}, and wirelessly-powered MECO \cite{wang2017joint,bi2017computation}. On the other hand, another line of research considers \emph{partially-synchronous} MECO, for which  the mobiles only share identical data-arrival time instants but may have different computation deadlines. For such systems, a set of offloading scheduling policies have been proposed to minimize the total mobile latency using techniques such as flow-shop queuing theory \cite{mao2017joint}, and joint scheduling and data partitioning \cite{ren2017latency}. 
 In addition, cooperative computing among mobiles was investigated in the recent work \cite{lee2017online,dinh2017offloading,you2017energy,ti2017computation} for reducing energy consumption and offloading latency via data partitioning and offloading scheduling techniques. {\color{black}{Specifically, the peer-to-peer offloading given the computation deadline was investigated in \cite{you2017energy} by using the ``string-pulling" approach.\footnote{{\color{black}{Compared with \cite{you2017energy}, the current work  considers heterogenous computation deadlines for different mobiles, which is more complex and thus cannot be directly solved using the ``string-puling" approach or traditional data-transmission techniques.}} }}} Note that in the above work, the assumption of synchronous or partially-synchronous MECO is unsuitable for many practical \emph{asynchronous} MECO systems that consist of mobiles with heterogeneous data-arrival time instants and deadlines. This motivates the current work that studies \emph{fully} asynchronous MECO systems.



Last, it is worth mentioning that in traditional communication systems without MECO, asynchronous packet transmission with individual latency constraints has been widely studied for designing offline and online scheduling policies\cite{zafer2005calculus,chen2007energy,rajan2004delay}.  {\color{black}{The above work only focuses on data transmissions following the first-come-first-serve rule. In contrast, for asynchronous MECO systems, the transmission techniques should be integrated with the joint radio-and-computational resource management, local computing, and interwound computation and transmission, which is the new theme of  this work.\footnote{{\color{black}{The current work differs from \cite{zafer2005calculus,chen2007energy,rajan2004delay} in the problem formulation and transformation, as well as providing new insights for asynchronous offloading.}}}}}

\subsection{Contributions}
\vspace{-5pt}
{\color{black}{
To the best of the authors' knowledge, this work was the first attempt on designing the energy-efficient offloading controller for practical asynchronous MECO systems with non-identical task-arrivals  and deadlines among mobiles. 
Compared with   synchronous MECO studied in most prior work, the current design eliminates  the overhead required  for network synchronization, and reduces offloading-and-computation latency. {\color{black}{Towards developing a framework for designing asynchronous offloading, the main  contributions of the work are twofold: 1) characterizing  the  structure of the optimal policy  that helps  simplify offloading-controller design and deepen the understanding of the technology,  and  2) proposing the approach for designing    practical offloading algorithms  via  decomposing  a complex  problem into low-complexity convex sub-problems. The specific  technical contributions and findings are summarized as follows.

\begin{itemize}

\item[1)] \emph{General arrival-deadline orders}: Consider the general case with arbitrary orders of data-arrival time instants and deadlines for different mobiles (see Fig.~\ref{Fig:System}). The design of offloading controller is formulated as an optimization problem under the criterion of minimum  total mobile-energy consumption and the constraints of  time-sharing and deadlines. An iterative solution method is proposed to iteratively optimize  data partitioning for individual  mobiles and multiuser time divisions. The computation  complexity is reduced by analyzing the policy structure for each iteration. 
The analysis reveals that  the optimal data-partitioning policy is characterized by a threshold-based  structure. Specifically,  each mobile should attempt to increase offloading or reduce it if the computation capacity of the mobile  or cloud server becomes a bottleneck as measured using corresponding derived thresholds. 

\item[2)] \emph{Identical arrival-deadline orders}: To gain more insights,
 consider the special case where the data-arrival time instants and deadlines of different mobiles follow the identical orders. 
 The optimization problem
   is decomposed   into two sequential problems, corresponding to  optimizing the \emph{scheduling order} and \emph{energy-efficient joint data partitioning and time division}  given the optimal order. Thereby, we show that without loss of optimality, the mobiles should be scheduled for offloading according to their data-arrival order. Leveraging this result, the original problem is simplified as the problem of    joint optimization of data partitioning and time division. Then the simplified problem is solved using the proposed  master-and-slave framework, where the slave problem optimizes data partitioning,  the master problem corresponds to the energy-efficient time division, and both are convex. Interestingly, it is discovered  that the optimal time-division policy attempts to equalize the differences in mobile computation capacities via  offloading time allocation to mobiles.

\item[3)] \emph{Reverse arrival-deadline orders}: For the same objective as the preceding task, we further consider another special case with  the reverse arrival-deadline orders, where a mobile with later data arrival must complete the computation earlier. The derived optimal scheduling order suggests two non-overlapping offloading intervals for each mobile. To obtain the optimal offloading durations given the optimal order, we propose a new and simple \emph{transformation-and-scheduling} approach.  Specifically, the transformation phase converts the original problem into the counterpart with identical arrival-deadline orders, allowing the use of the previous  solution approach. Then given the scheduling order, individual offloading intervals are computed in the the scheduling phase.
\end{itemize}}}}
{\color{black}{The differences between this paper and its conference version \cite{csyou2018asynMEC} are as follows. First, this paper considers the finite computation capacities at the mobiles and edge cloud, while infinite computation capacities are assumed in \cite{csyou2018asynMEC}. Second, several useful discussions are added in this paper to demonstrate the versatility of proposed algorithms. Last, the paper studies the resource management for the case of reverse arrival-deadline orders, which is not addressed in \cite{csyou2018asynMEC}.}}

\section{System Model}\label{Sec:Sys}

Consider a multiuser MECO system (see Fig.~\ref{Fig:System}), comprising one single-antenna  BS  connected to an edge cloud and $K$ single-antenna mobiles, denoted by a set $\mathcal{K}=\{1, 2, \cdots, K\}$. 
Each mobile has one-shot input-data arrival at a random time instant and is required to complete the computation before a given deadline. We consider asynchronous computation offloading, where the data-arrival time instants and deadlines vary for different mobiles. The  input data is partitioned into two parts for parallel computation: one at the mobile's local CPU and the other offloaded to the BS.\footnote{\color{black}{For tractability, we assume that the input data can be arbitrarily partitioned following the literature (see e.g., \cite{you2015energyJSAC}). This is in fact the case for certain applications such as Gzip compression and feature extraction.}}  {\color{black}{In the message-passing phase prior to computation offloading, each mobile feeds back to the BS its state parameters, including the estimated channel gain, data-arrival time instant and deadline (acquired by  CPU profiling or CPU-utilization prediction techniques \cite{dinda1999evaluation,xu2017enabling}). Using the information, the BS determines the energy-efficient  resource-management policy for controlling the mobiles' offloaded bits and durations, and then broadcasts  the control policy to mobiles. 
 }} 
\vspace{-5pt}
\subsection{Model of Input-Data Arrivals}
The asynchronous data arrivals for the mobiles are modeled as follows. As shown in Fig.~\ref{Fig:System}, each mobile, say mobile $k$, needs to complete a computation task with $L_k$-bit input data within the time interval $\l[T_k^{(a)}, T_k^{(d)}\r]$, where $T_k^{(a)}$ is the data-arrival time instant and $T_k^{(d)}$ is the computation deadline. The required computation latency for mobile $k$, denoted by $T_k$, is thus given by $T_k=T_k^{(d)}-T_k^{(a)}$, in second (s). Without loss of generality, assume that $T_1^{(a)}\le T_2^{(a)}\le \cdots \le T_K^{(a)}$ and $T_1^{(a)}=0$.\footnote{We assume that $T_k^{(d)}>T_{k+1}^{(a)}$ for $k=1, 2, \cdots, K-1$, such that the computing intervals of each mobile always overlaps with that of others (see Fig.~\ref{Fig:System}). Otherwise, the total duration can be decoupled into several non-overlapping durations.} To facilitate the exposition  in the sequel, we define two useful sets as below.

\begin{definition}[Epoch-Set, User-Set]\label{Def:EpockUser}\emph{Let $\{s_n\}$ with $n=0, 1,\cdots, N=2K-1,$ denote a sequence of ordered time instants and $\mathbf{\Pi}$ the permutation matrix given by  $$[s_0, s_1, \cdots, s_N]^{T}=\mathbf{\Pi}\times [T_1^{(a)}, T_2^{(a)}, \cdots T_K^{(a)}, T_1^{(d)}, T_2^{(d)}, \cdots, T_K^{(d)}]^T,$$ such that $s_0\le s_1\le \cdots\le s_N$, and $s_0=T_1^{(a)}$.  The time interval between two consecutive time instants is called an \emph{epoch} with length $\tau_n\overset{\triangle}{=}s_n-s_{n-1}$ for $n=1, 2,\cdots N$. For each mobile, say mobile $k$, let $A_k$ denote its \emph{epoch-set} which specifies the indexes of epochs that constitute the computing interval of mobile $k$. For each epoch, say epoch $n$, define the \emph{user-set}  $B_n$ as the indexes of mobiles whose computing intervals cover epoch $n$. 
}
\end{definition}
For an example shown in Fig.~\ref{Fig:System}, the epoch set for mobile $1$ is $A_1=\{1,2,3\}$, and the user-set for epoch $2$ is $B_2=\{1,2\}$. {\color{black}{If given $T_1^{(a)}=0, T_1^{(d)}=5, T_2^{(a)}=3, T_2^{(d)}=7, T_3^{(a)}=4$, and $T_3^{(d)}=6$, $\mathbf{\Pi}$ can be constructed as $\mathbf{\Pi}=[\bf{e}_1, \bf{e}_3, \bf{e}_5, \bf{e}_2, \bf{e}_6, \bf{e}_4]^{T}$ where the $6\times1$ vector $\bf{e}_n$ is the $n$-th column of the identity matrix ${\bf{I}}$.}}

\vspace{-5pt}
\subsection{Models of Local Computing and Computation Offloading}
Let $\ell_{k,n}$ denote the offloaded bits of mobile $k$ during epoch $n$. To finish the computation before the deadline, the remaining $(L_k-\sum_{n\in A_k} \ell_{k,n})$-bit data is computed by the mobile's CPU. The models of local computing and computation offloading are described as follows.

\subsubsection{Local Computing}
Based on the model in \cite{xu2015mec}, let $C_k$ denote the number of CPU cycles required for computing $1$-bit data for mobile $k$, which may be different for different mobiles depending on their specific computing-task complexities. During the computing duration $T_k$, since operating at a constant CPU-cycle frequency is  most energy-efficient for local computing  \cite{mao2016dynamic}, the  CPU-cycle frequency for mobile $k$ is chosen as $f_k=C_k(L_k-\sum_{n\in A_k} \ell_{k,n})/T_k$.   Following the model in \cite{burd:CpuEnergy:1996}, under the assumption of  low CPU voltage, the energy consumption for each CPU cycle can be modeled by $E_{\text{cyc}, k}(f_k)=\gamma f_k^2$, where $\gamma$ is a constant determined by the circuits. Then the local-computing energy consumption for mobile $k$, denoted by $E_{\text{loc},k}$, is obtained as:
\begin{align*}
\text{(Local-computing energy consumption)}\quad  E_{\text{loc},k}=\dfrac{\gamma C_k^3\l(L_k-\sum_{n\in A_k} \ell_{k,n}\r)^3}{T_k^2}.
\end{align*}
{\color{black}{Let $F_k$ denote the maximum CPU frequency of mobile $k$. Then we have $C_k(L_k-\sum_{n\in A_k} \ell_{k,n})/T_k\le F_k$. As a result, the offloaded data size of mobile $k$ is lower-bounded as $\sum_{n\in A_k} \ell_{k,n}\ge R_k^{(\rm{min})}$, where $R_k^{(\rm{min})}=\max\{L_k-T_kF_k/C_k, 0\}$.}}

\subsubsection{Computation Offloading}\label{Sec:Off}
 For each mobile, computation offloading comprises three sequential phases: 1) offloading data from the mobile to the edge cloud, 2) computation by the edge cloud, and 3) downloading of computation results from the edge cloud to the mobile. {\color{black}{Assume that the edge cloud assigns an individual \emph{virtual machine} (VM) for each mobile using VM multiplexing and consolidation techniques that allow for multi-task parallel computation \cite{ahmad2015survey}. Based on the model in \cite{you2016energy}, the \emph{finite} VM computation capacity for each mobile can be reflected by upper-bounding the number of offloaded CPU cycles, denoted by $D_k$, for which the required computation time remains \emph{negligible} compared with the total computation latency $T_k$.  Mathematically,  it enforces that $C_k (\sum_{n\in A_k} \ell_{k,n})\le D_k$.}}
 {\color{black}{Moreover, assuming relatively small sizes of computation results for   applications (such as face recognition, object detection in video, and   online chess game) and high transmission power at the BS, downloading is much faster than offloading and consumes negligible mobile energy.{\footnote{\color{black}{For data-intensive applications such as virtual/augmented reality, the energy consumption and latency for result downloading is non-negligible. In these cases, we expect that the current framework for offloading can be modified and  applied to designing  asynchronous downloading control as well.}}}}} Under these conditions, the second and third phases are assumed to have negligible durations compared with the first phase. Assume that the mobiles access the cloud based on TDMA. Specifically, for  each epoch, say epoch $n$, the mobiles belonging to the user-set $B_n$ time-share the epoch duration $\tau_n$. For these mobiles, let $t_{k,n}$ denote the allocated offloading duration for mobile $k$,  where $t_{k,n} =0$ corresponds to no offloading. For the case of offloading ($t_{k,n} > 0$), let $g_k$ denote the channel power gain between mobile $k$ and the BS, which is assumed to be constant during the computation offloading for each mobile.   
{\color{black}{Based on a widely-used empirical model in \cite{zafer2007delay,lee2009delay,zhang:MobileMmodel:2013,ko2017live}}}, the transmission power, denoted by $p_{t,n}$, can be modeled by a \emph{monomial} function with respect to the achievable transmission rate (in bits/s) $r_{k,n}=\ell_{k,n}/t_{k,n}$: 
\begin{equation}
\text{(Monomial offloading power)}\quad P_{k,n}=\frac{\lambda (r_{k,n})^m}{g_k},
\end{equation}
where $\lambda$ denotes the energy coefficient incorporating the effects of bandwidth and noise power, and $m>1$ is the monomial order determined by the adopted coding scheme. {\color{black}{Though this assumption may restrict the generality of the problem, it  leads to simple solutions in (semi-) closed forms as shown in the sequel and provides useful insights for practical implementation. Moreover, it provides a good approximation for the transmission power of practical transmission schemes.}} For example, considering the coding scheme for the targeted bit error probability less than $10^{-6}$ \cite{neely2005dynamic}, Fig.~\ref{Fig:Monomial_test} gives the normalized signal power per symbol versus the rate, where the monomial order of $(m=3)$ can fairly approximate the transmission power.\footnote{{\color{black}{In practice, the value of $m$ can be determined by curve-fitting using experimental data. Note that it is possible to achieve better curve-fitting performance by using the polynomial function, for which the proposed iterative design in the sequel can be extended to solve the corresponding convex optimization problem with key procedures remaining largely unchanged.}}}  Thus, the offloading energy consumption can be modeled by   the following monomial function with respect to $\ell_{k,n}$ and $t_{k,n}$:
\begin{equation}\label{Eq:OffEgy}
\vspace{-5pt}
\text{(Monomial offloading energy consumption)}\quad E_{{\rm{off}},k,n}=P_{k,n} t_{k,n}= \frac{\lambda (\ell_{k,n})^m}{g_k(t_{k,n}) ^{m-1}}. 
\end{equation}
Note that if $t_{k,n} = 0$, we have $\ell_{k,n}=r_{k,n} t_{k,n}=0$ and thus $E_{{\rm{off}},k,n}=0$. The total energy consumption of mobile $k$ for transmitting  the offloaded input data, denoted by $E_{{\rm{off}},k}$, is given by: $E_{{\rm{off}},k}=\sum_{n\in A_k} E_{{\rm{off}},k,n}$.  
\begin{figure}[t!]
  \centering
  \subfigure{
  \begin{minipage}{5cm}
  \centering
  \includegraphics[width=5cm]{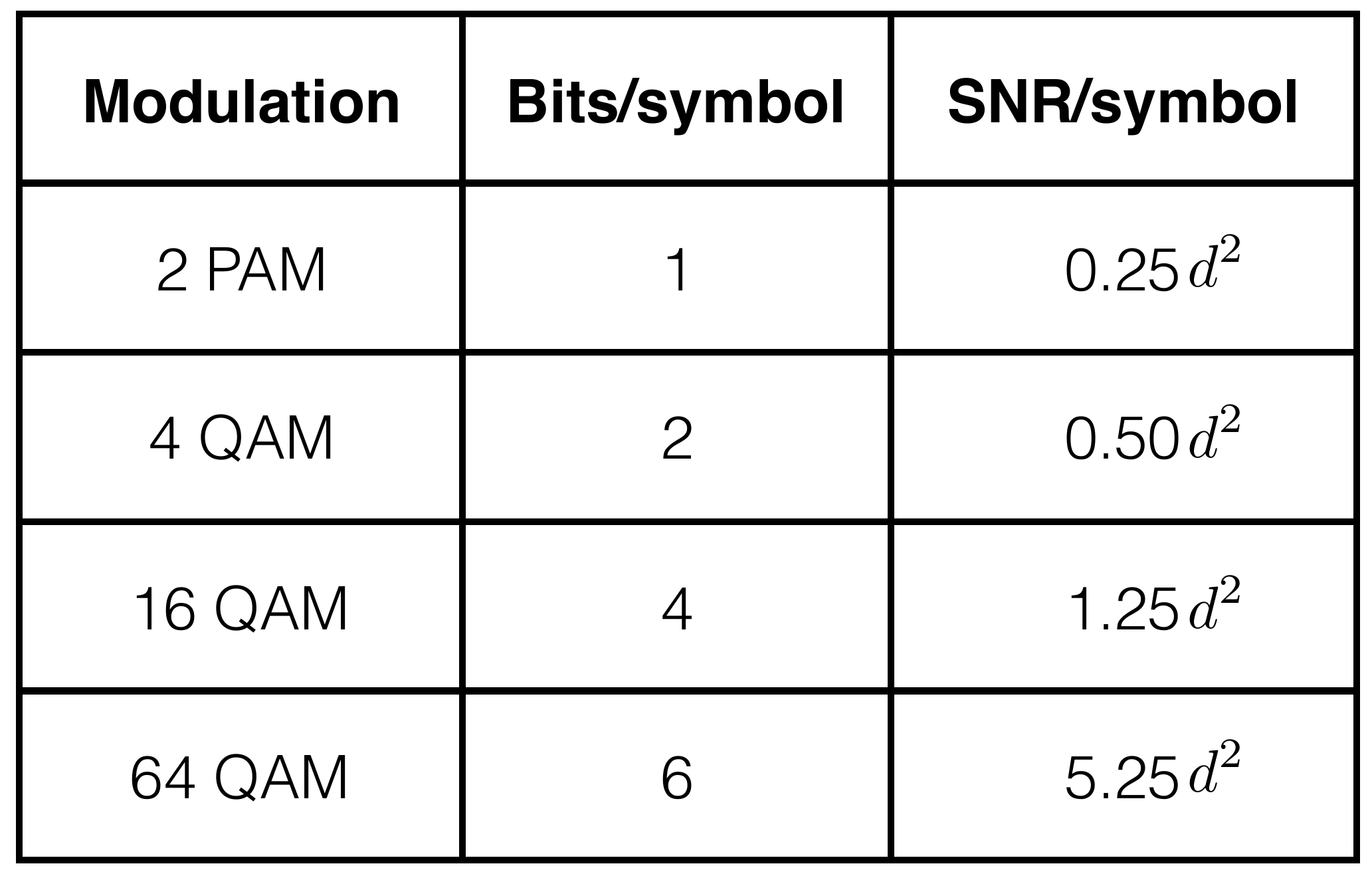}
  \end{minipage}
  }
  \subfigure{
  \begin{minipage}{4.8cm}
  \centering
  \includegraphics[width=4.8cm]{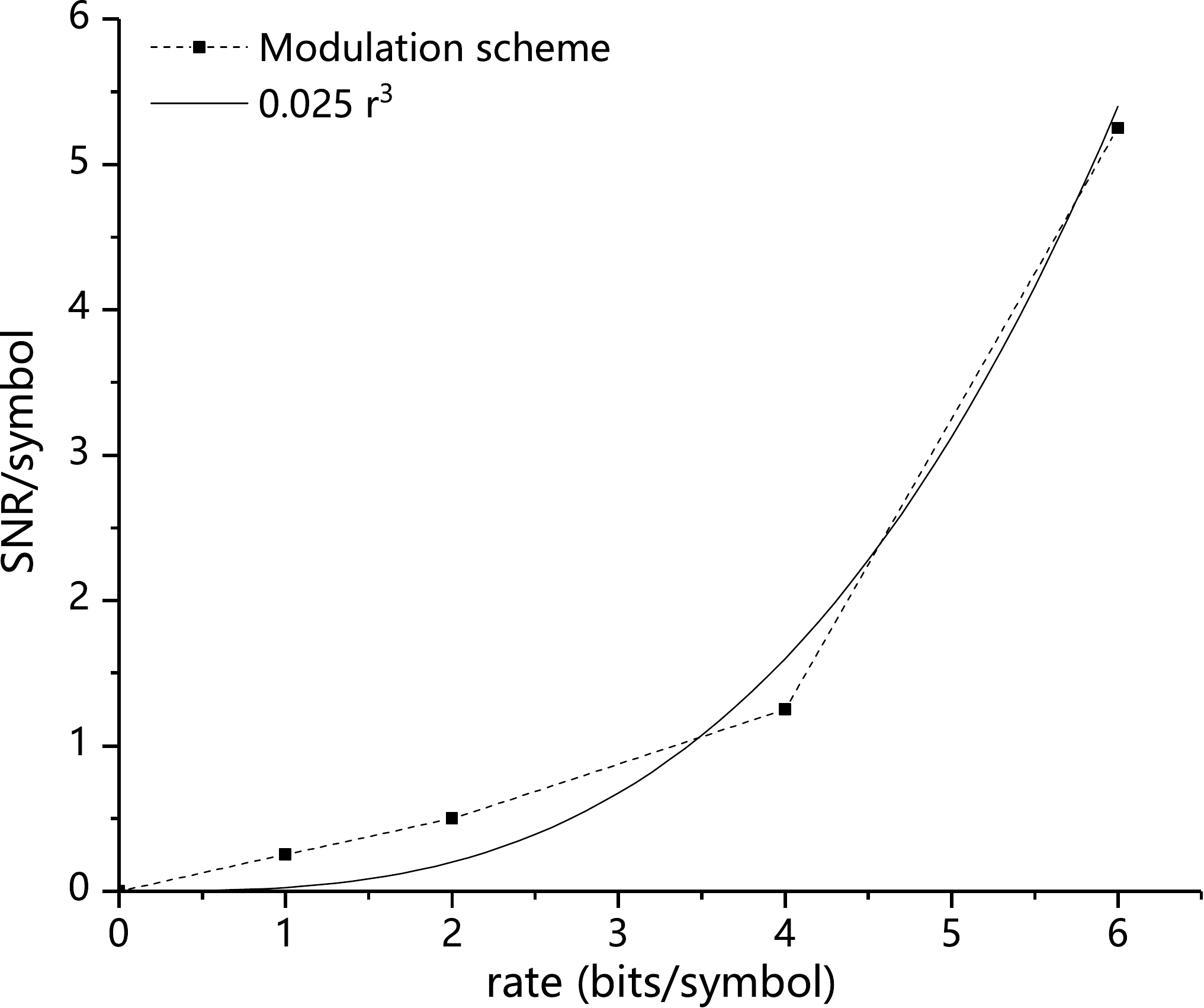}
  \end{minipage}
  }
  \caption{Modulation scheme given in the table is considered in \cite{neely2005dynamic}, where SNR is short for signal-to-noise ratio and $d$ represents the minimum distance between signal points. The corresponding plot shows $0.025r^3$ to the scaled piecewise linear power-rate curve.}\label{Fig:Monomial_test}
 \vspace{-8pt}
\end{figure}

\vspace{-15pt}
\section{Problem Formulation}\label{Sec:ProFor}
In this section, the energy-efficient asynchronous MECO resource management is formulated as an optimization problem that jointly optimizes the data partitioning and  time divisions for the mobiles.  The objective is to minimize the  total mobile-energy consumption: $\sum_{k=1}^K  (E_{\text{off},k}+E_{\text{loc},k})$. For each epoch, the multiuser offloading should satisfy the time-sharing constraint: 
\begin{equation}\label{Eq:TimeShare}
\vspace{-5pt}
\text{(Time-sharing constraint)}\quad \sum_{k\in B_n} t_{k,n}\le \tau_n, \quad \forall n.
\end{equation} 
For each user, the total offloaded data size and computation are constrained by: 
\begin{align}
\text{(Data constraint)}\quad &\sum_{n\in A_k} \ell_{k,n}\le L_k, \quad\forall k,\\
{\color{black}{\text{(Local computation capacity constraint)}}}\quad & {\color{black}{\sum_{n\in A_k}\ell_{k,n} \ge R_k^{\rm{min}},\quad\forall k,}}\\
{\color{black}{\text{(VM computation capacity constraint)}}}\quad &{\color{black}{C_k(\sum_{n\in A_k} \ell_{k,n})\le D_k, \quad\forall k}}.
\end{align}
Note that the deadline constraint for each mobile is enforced by setting the local-computing data size as $(L_k-\sum_{n\in A_k} \ell_{k,n})$-bits. Under these constraints, the optimization problem is readily formulated as: 
{\color{black}{\begin{equation}\tag{$\textbf{P1}$} 
\begin{aligned}
\min_ {\{\ell_{k,n}\ge0, t_{k,n}\ge0\} }   ~ &\sum_{k=1}^{K}  \l[\l(\sum_{n\in A_k} \frac{\lambda(\ell_{k,n})^m}{g_k(t_{k,n}) ^{m-1}}\r) + \dfrac{\gamma C_k^3(L_k-\sum_{n\in A_k} \ell_{k,n})^3}{T_k^2} \r] \\
\text{s.t.}\qquad 
&R_k^{(\min)} \le \sum_{n\in A_k} \ell_{k,n}\le R_k^{(\max)},  &&\forall k,\\
& \sum_{k\in B_n} t_{k,n} \le \tau_n,  && \forall n,
\end{aligned}
\end{equation}
where $R_k^{(\max)}=\min\{L_k, D_k/C_k\}$.}}
{\color{black}{One can observe that Problem P1 is feasible if and only if $R_k^{\min}\le R_k^{\max}$, which is equivalent to $L_k-\frac{T_kF_k}{C_k}\le \frac{D_k}{C_k}$.}}
 Next, note that the variables $\{\ell_{k,n}\}$ and $\{t_{k,n}\}$ are coupled in the objective function. To overcome this difficulty, one important property of Problem P1 is provided in the following lemma, which can be proved
 in Appendix~\ref{App:P1Conv}.
\begin{lemma}\label{Lem:P1Conv}\emph{Problem P1 is a convex optimization problem.}
\end{lemma}
Thus, Problem P1 can be directly solved by the Lagrange method that involves the primal and dual problem optimizations \cite{Boyd2006convex}. {\color{black}{This method, however, cannot provide useful insights on the structure of the optimal policy, since it requires the joint optimization for the data partitioning and time division that have no closed form.}} To address this issue, 
 in the following sections, we first study the optimal resource-management policy for the \emph{general} case where deadlines of mobiles are arbitrary  (e.g., $T_3^{(d)}\le T_5^{(d)}\le \cdots \le T_2^{(d)}$) by using  the \emph{block coordinate decent} (BCD) optimization method \cite{hong2016unified}. Subsequently, we derive more insightful structures of the optimal policy for two special cases, namely asynchronous  MECO with the \emph{identical} and \emph{reverse} arrival-deadline orders. Recall that for the data-arrival order, we have $T_1^{(a)}\le T_2^{(a)}\le \cdots \le T_K^{(a)}$ without loss of generality. The so-called identical and reverse arrival-deadline orders refer to the cases where it satisfies $T_1^{(d)}\le T_2^{(d)}\le \cdots \le T_K^{(d)}$ and  $T_1^{(d)}\ge T_2^{(d)}\ge \cdots \ge T_K^{(d)}$, respectively, as illustrated in Fig.~\ref{Fig:IdenRever}.
\begin{figure}[t]
\begin{center}
\includegraphics[height=4.5cm]{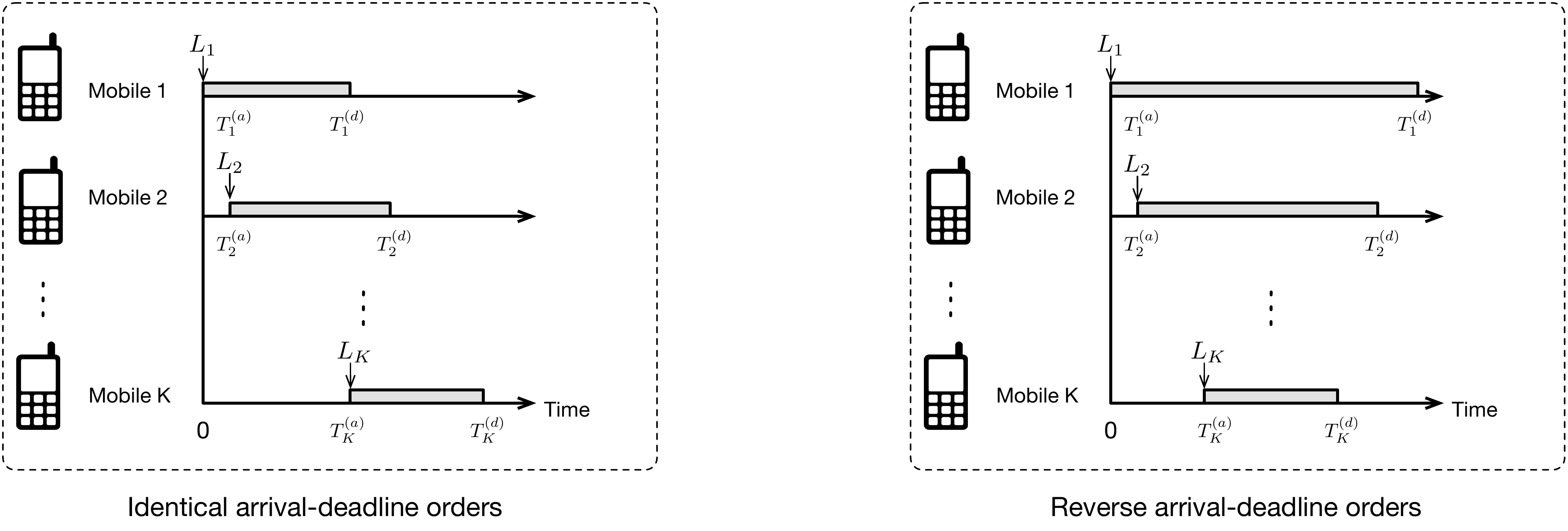}
\caption{Illustration for asynchronous MECO systems with the identical and reverse arrival-deadline orders.}
\label{Fig:IdenRever}
\end{center}
\vspace{-15pt}
\end{figure}

\section{Optimal Resource Management with  General Arrival-Deadline Orders}\label{Sys:Gen}
This section considers the asynchronous  MECO with general arrival-deadline orders and designs the energy-efficient resource-management policy.
 To characterize the {\color{black}{structures of the optimal policy}}, we propose an iterative algorithm for solving Problem P1 by applying the BCD method. Specifically, given any offloading durations for all the mobiles  $\{t_{k,n}\}$, we optimize the offloaded data sizes $\{\ell_{k,n}\}$ for each mobile, corresponding to energy-efficient \emph{data partitioning}. On the other hand, the offloading durations of  the mobiles, $\{t_{k,n}\}$, are optimized given any offloaded data sizes $\{\ell_{k,n}\}$, referred to as  energy-efficient \emph{time division}.

\vspace{-5pt}
\subsection{Energy-Efficient Data Partitioning}
This subsection aims at finding the optimal offloaded data sizes $\{\ell_{k,n}\}$ for the mobiles, given any feasible offloading time divisions $\{t_{k,n}\}$. For each mobile $k$, let $\bar{A}_k$ denote its offloading epoch set comprising the epoch indexes for which $t_{n,k}>0$. Mathematically, $\bar{A}_k=\{n\in A_k, | t_{n,k}>0\}$. Then it can be easily observed that Problem P1 reduces to $K$ parallel sub-problems, each corresponding to one mobile as:
\begin{equation}\tag{$\textbf{P2}$} 
\begin{aligned}
\min_ {\{\ell_{k,n}\ge0\} }   ~ &\sum_{n\in \bar{A}_k} \frac{\lambda(\ell_{k,n})^m}{g_k(t_{k,n}) ^{m-1}} + \dfrac{\gamma C_k^3(L_k-\sum_{n\in \bar{A}_k} \ell_{k,n})^3}{T_k^2}
\quad \text{s.t.} 
& R_k^{(\min)} \le \sum_{n\in A_k} \ell_{k,n}\le R_k^{(\max)}.
\end{aligned}
\end{equation}
Problem P2 can be easily proved to be a convex optimization problem. Applying the Lagrange method leads to the optimal data-partitioning policy as follows, which is proved in Appendix~\ref{App:GeneDataParti}.

{\color{black}{\begin{proposition}[Energy-Efficient Data Partitioning]\label{Prop:GeneDataParti}\emph{
For each mobile, say mobile $k$, given the offloading time divisions $\{t_{k,n}\}$, the optimal data-partitioning policy for different epochs for solving Problem P2, denoted by $\{\ell_{k,n}^*\}$, is given by
\begin{equation}\label{Eq:GeneOffBits}
\ell_{k,n}^*=
\begin{cases}
h(\xi_k^*), & R_k^{(\min)}\le \sum_{n\in \bar{A}_k} h(\xi_k^*)\le R_k^{\max}, 
\cr \dfrac{t_{k,n}}{\sum_{n\in \bar{A}_k} t_{k,n}} R_k^{(\min)}, & \sum_{n\in \bar{A}_k} h(\xi_k^*)<R_k^{(\min)},
\cr \dfrac{t_{k,n}}{\sum_{n\in \bar{A}_k} t_{k,n}} R_k^{(\max)}, & \sum_{n\in \bar{A}_k} h(\xi_k^*)>R_k^{(\max)},
\end{cases}
\end{equation}
for $\forall n\in \bar{A}_k$, where 
\begin{equation}\label{Eq:h_xi}
h(\xi_k^*)=\l({\dfrac{3b_k \xi_k^*}{m a_k T_k^{2}}}\r)^{\frac{1}{m-1}} t_{k,n},
\end{equation} $a_k=\dfrac{\lambda}{g_k}$, $b_k=\gamma C_k^3$, and $\xi_k^*>0$ is the solution to $U_k(\xi_k)=0$, with
\vspace{-10pt}
\begin{equation}\label{Eq:U_k}
U_k(\xi_k)\overset{\triangle}{=}\l(L_k-\sum_{n\in \bar{A}_k} h(\xi_k)\r)^2-\xi_k.
\end{equation} 
 }
\end{proposition}}}
Proposition~\ref{Prop:GeneDataParti} shows that the optimal offloaded data sizes, $\{\ell_{k,n}^*\}$, are determined by a single parameter $\xi_k^*$. Specifically, as $U_k(0)=L_k^2>0$, $U_k(\hat{\xi}_k)=-\hat{\xi}_k<0$ where
 $\hat{\xi}_k= \dfrac{m a_k T_k^{2} (R_k^{(\max)})^{m-1}}{{3b_k \l(\sum_{n\in A_k} t_{k,n}\r)^{m-1}}},$
 and  $U_k(\xi_k)$ defined in \eqref{Eq:U_k} monotonically decreases with $\xi_k$,  $\xi_k^*$ and thus $\{\ell_{k,n}^*\}$ can be uniquely determined and   efficiently computed using the bisection-search algorithm  \cite{Boyd2006convex}. {\color{black}{In addition, it can be proved by contradiction that for the case of $R_k^{(\min)}\le \sum_{n\in \bar{A}_k} h(\xi_k^*)\le R_k^{\max}$, the optimal offloaded data size in each epoch, $\ell_{k,n}^*$, is \emph{monotonically-increasing} with $g_k$, $C_k$ and $t_{k,n}$, and \emph{monotonically-decreasing} with $T_k$ and $m$. This is consistent with the intuition that it is desirable to offload more bits as the channel condition improves, the local-computing complexity increases, the allocated offloading time duration increases, or the computation deadline requirement becomes more stringent. Moreover, when the monomial order increases (e.g., when the offloading wireless transmission targets for a lower error probability), it is more energy-efficient to reduce the offloaded data size since the required transmission power increases with $m$.}}
 
 \begin{remark}[Identical Offloading Rates]\emph{It can be inferred from Proposition~\ref{Prop:GeneDataParti} that given the optimal time divisions $\{t_{k,n}\}$,  for each mobile, the optimal offloading rates $r_{k,n}^*=\frac{\ell_{k,n}^*}{t_{k,n}^*}$ in different epochs are identical. This is expected, since for each mobile, the channel power gain, bandwidth and noise power are the same in different epochs.}
\end{remark}
{\color{black}{To further characterize the effects of offloading duration and computation capacities of the mobile and cloud on the data-partitioning policy, we define an auxiliary function $\phi_k(\{t_{k,n} | n\in \bar{A}_k\})$ for each mobile $k$, denoted as $\phi_k$ for simplicity, as the root of the following equation with respect to $x$.
\begin{equation}\label{Eq:Phi}
\frac{(L_k-x)^{m-1}}{x^2}=\frac{3b_k}{m a_k T_k^2} \l(\sum\nolimits_{n\in \bar{A}_k} t_{k,n}\r)^{m-1}.
\end{equation}
Two useful properties of $\phi_k$ can be easily derived: 1) $0<\phi_k<L_k$, and 2) $\phi_k$ is monotonically decreasing with the total offloading duration $\sum_{n\in \bar{A}_k} t_{k,n}$. Then the optimal data-partitioning policy in Proposition~\ref{Prop:GeneDataParti} can be restated as follows, which is proved in Appendix~\ref{App:GeneDataPartiRe}.
\begin{corollary}\label{Cor:GeneDataPartiRe}\emph{
For each mobile, say mobile $k$, given the offloading time divisions $\{t_{k,n}\}$, the optimal data-partitioning policy in Proposition~\ref{Prop:GeneDataParti} can be re-expressed as 
{\color{black}{
\begin{equation}\label{Eq:GeneOffBits}
\ell_{k,n}^*=
\begin{cases}
h(\xi_k^*), & F_k\ge\dfrac{C_k \phi_k}{T_k}~\text{and}~D_k\ge C_k(L_k-\phi_k),
\cr \dfrac{t_{k,n}}{\sum_{n\in \bar{A}_k} t_{k,n}} \l(L_k-\dfrac{F_k T_k }{C_k}\r), &F_k<\dfrac{C_k \phi_k}{T_k}~\text{and}~D_k\ge C_k(L_k-\phi_k),
\cr \dfrac{t_{k,n}}{\sum_{n\in \bar{A}_k} t_{k,n}} \l(\dfrac{D_k}{C_k}\r), &
F_k\ge\dfrac{C_k \phi_k}{T_k}~\text{and}~D_k< C_k(L_k-\phi_k),
\end{cases}
\end{equation}
for $\forall n\in \bar{A}_k$, where $h(\xi_k^*)$ is defined in \eqref{Eq:h_xi}.
 }}}
\end{corollary}
Corollary~\ref{Cor:GeneDataPartiRe} shows that each mobile $k$ should perform the \emph{mobile-constrained minimum} or \emph{cloud-constrained maximum} computation offloading (with the total offloaded data sizes being $L_k-\frac{F_k T_k }{C_k}$ and $\frac{D_k}{C_k}$, respectively), if the mobile or VM server becomes a bottleneck with insufficient computation capacities less than the given thresholds, respectively. It is worth mentioning that if both the mobile and VM have insufficient capacities, computing the input-data by the deadline is infeasible. Moreover, it can be observed that as the total offloading duration grows, $\frac{C_k \phi_k}{T_k}$ decreases and $C_k(L_k-\phi_k)$ increases, meaning that the mobile tends to offload more data provisioned with a longer offloading duration.}}

\vspace{-7pt}
\subsection{Energy-Efficient Time Division}\label{Sec:GeneTimeDiv} For given offloaded data sizes $\{\ell_{k,n}\}$, this subsection focuses on optimizing the time-division policy, $\{t_{k,n}\}$, in all epochs to minimize the total mobile-energy consumption.  For each epoch $n$, let $\bar{B}_n$ denote the offloading user-set comprising the mobile indexes for which $\ell_{n,k}>0$. Mathematically, $\bar{B}_n=\{k\in B_n, | \ell_{n,k}>0\}$. Since the time-sharing constraints can be decoupled for different epochs, Problem P1  reduces to solving the following $N$ parallel sub-problems:
\begin{equation}\tag{$\textbf{P3}$}
\begin{aligned}
\min_ {\{t_{k,n}\ge0\} }   &~ \sum_{k\in \bar{B}_n}  \frac{\lambda(\ell_{k,n})^m}{g_k(t_{k,n}) ^{m-1}}
\qquad \text{s.t.}
& \sum_{k\in \bar{B}_n} t_{k,n} \le \tau_n, ~~\forall n.
\end{aligned}
\end{equation}
Problem P3 is a convex optimization problem and its optimal solution can be easily derived by using the Lagrange method, which is given in the following proposition.
\begin{proposition}[Energy-Efficient Time Division]\label{Pro:GeneTimeDiv}\emph{For each epoch, say epoch $n$, given any offloaded data sizes $\{\ell_{k,n}>0\}$, the optimal time-division policy for different mobiles for solving Problem P3, denoted by $\{t_{k,n}^*\}$, is given by 
\begin{equation}
t_{k,n}^*=\frac{\eta_{k,n}}{\sum_{\forall k\in \bar{B}_n}\eta_{k,n}} \tau_n, \qquad \forall k\in \bar{B}_n,
\end{equation}
where $\eta_{k,n}=\l(\dfrac{(m-1) \lambda}{g_k}\r)^{\frac{1}{m}}\ell_{k,n}$.
 }
\end{proposition}
Proposition~\ref{Pro:GeneTimeDiv} shows that the optimal offloading duration for each mobile is \emph{proportional} to the epoch duration by a proportional ratio $\dfrac{\eta_{k,n}}{\sum_{\forall k\in \bar{B}_n}\eta_{k,n}}$, which is determined by the offloaded data size and channel gain. Specifically, to minimize the total mobile-energy consumption in each epoch, the mobile with a larger offloaded data size and poorer  channel should be allocated with a longer offloading duration.

Last, based on the results obtained in these two subsections, the optimal solution to Problem P1 can be efficiently computed by the proposed iterative algorithm using the BCD method, which is summarized in Algorithm~\ref{Alg:GeneBCD}. Since Problem P1 is jointly convex with respect to the data partitioning $\{\ell_{k,n}\}$ and time divisions $\{t_{k,n}\}$, iteratively solving Problem P2 and P3 can guarantee the convergence to the optimal solution to Problem P1. 
{\color{black}{\begin{remark}[Low-Complexity Algorithm]\emph{Given offloading time-divisions, the computation complexity for the optimal data partitioning is up to $\mathcal{O}(K\log(1/\epsilon))$, where  $\log(1/\epsilon)$ characterizes the complexity order for the one-dimensional search. Given offloaded data sizes, the optimal time-division policy has the complexity order of $\mathcal{O}(N)$ owing to the closed-form expression. Thus, the total computation complexity for the proposed BCD algorithm is $\mathcal{O}(K\log^2(1/\epsilon)\!+\!N \log(1/\epsilon))$ accounting for the iterative procedures. Simulation results in the sequel show that the proposed method can greatly reduce the computation complexity, especially for larger number of mobiles and epochs compared with the general convex optimization solvers, e.g. CVX, which is based on the standard interior-point method that has the complexity order of $\mathcal{O}((NK)^{3.5}\log(1/\epsilon))$  \cite{ben2001lectures}.}
\end{remark}}}

\begin{algorithm}[t!]
  \caption{The Proposed Block Coordinate Descent Method for Problem P1}
  \label{Alg:GeneBCD}
  \begin{itemize}
\item{\textbf{Step 1} [Initialize]: Let  $t_{k,n}^{(0)}=\tau_n/{|B_n|}, \forall n, k$; $\epsilon>0$, and $r =0$.
}
\item{\textbf{Step 2}  [Block coordinate descent method]: \emph{Repeat}\\
(1) Given $\l\{t_{k,n}^{(r)}\r\}$, compute the optimal data-partitioning policy $\l\{\ell_{k,n}^{(r+1)}\r\}$ as in Proposition~\ref{Prop:GeneDataParti}.\\
(2) Given $\l\{\ell_{k,n}^{(r+1)}\r\}$, compute the optimal time-division policy $\l\{t_{k,n}^{(r+1)}\r\}$ as in Proposition~\ref{Pro:GeneTimeDiv}.\\
(3) Update $r=r+1$.\\
\emph{Until}: The fractional decrease of the objective value of Problem P1 is below a threshold $\epsilon$.}
\end{itemize}
  \end{algorithm}
 
     {\color{black}{\subsection{Extension: Asynchronous  MECO Based on Exponential Offloading Energy-Consumption Model}\label{Sys:ExtExp}
In this subsection, the solution approach developed in the preceding subsections is extended to the case with the exponential offloading energy-consumption model. Specifically, based on  Shannon's equation,  the achievable rate $r_{k,n}$ can be expressed as $r_{k,n}=B \log_{2}\l(1+\dfrac{p_{k,n} g_k}{N_0}\r)$ where $B$ denotes the bandwidth, and $N_0$ the noise power. Since constant-rate transmission is the most energy-efficient transmission policy \cite{you2015energyJSAC}, it follows that the energy consumption for offloading $\ell_{k,n}$-bit  data with duration  $t_{k,n}$ is given by 
\begin{equation}
\text{(Exponential offloading energy consumption)}\quad E_{{\rm{off}},k,n}=\frac{t_{k,n}}{g_k}\psi\!\l(\frac{\ell_{k,n}}{t_{k,n}}\r),
\end{equation} where the function $\psi(x)$ is defined as $\psi(x)=N_0 (2^{\frac{x}{B}}-1)$. Based on this model, Problem P1 is modified by replacing the objective function with the following and the resulting new problem is denoted as Problem P4.
 \begin{equation}
 \min_ {\{\ell_{k,n}\ge0, t_{k,n}\ge0\} }   ~ \sum_{k=1}^{K}  \l[\l(\sum_{n\in A_k} \frac{t_{k,n}}{g_k}\psi\!\l(\frac{\ell_{k,n}}{t_{k,n}}\r)\r) + \dfrac{\gamma C_k^3(L_k-\sum_{n\in A_k} \ell_{k,n})^3}{T_k^2} \r]. 
 \end{equation} 
 By following the similar procedure as for deriving Lemma~\ref{Lem:P1Conv}, it can be shown that Problem P4 is a convex optimization problem. To characterize its optimal policy structure, we apply the BCD method to derive the energy-efficient data-partitioning and time-division policies as detailed in the following.
\subsubsection{Energy-Efficient Data Partitioning} For any given offloading division $\{t_{k,n}\}$, Problem P4 reduces to $K$ parallel sub-problems:
 \begin{equation}\tag{$\textbf{P5}$} 
\begin{aligned}
\min_ {\{\ell_{k,n}\ge0 \}}   ~ &\l(\sum_{n\in \bar{A}_k} \dfrac{t_{k,n}}{g_k}\psi\!\l(\dfrac{\ell_{k,n}}{t_{k,n}}\r)\r) + \dfrac{\gamma C_k^3(L_k-\sum_{n\in \bar{A}_k \ell_{k,n}})^3}{T_k^2} 
\quad \text{s.t.} 
&  R_k^{\min}\sum_{n\in \bar{A}_k} \ell_{k,n}\le R_k^{\max},
\end{aligned}
\end{equation}
where $\bar{A}_k$ is similarly defined as in Problem P2.
Problem P5 is a convex optimization problem. Directly applying Lagrange methods yields the optimal solution as below.
\begin{proposition}\label{Prop:ExpoData}\emph{Consider asynchronous MECO based on the exponential offloading energy-consumption model. For any given offloading time division $\{t_{k,n}\}$, the optimal offloading data size for each mobile is given by 
\begin{equation}
\ell_{k,n}^*=
\begin{cases}
\widetilde{h}(\xi_k^*), & R_k^{(\min)}\le \sum_{n\in \bar{A}_k} \widetilde{h}(\xi_k^*)\le R_k^{\max}, 
\cr \dfrac{t_{k,n}}{\sum_{n\in \bar{A}_k} t_{k,n}} R_k^{(\min)}, & \sum_{n\in \bar{A}_k} \widetilde{h}(\xi_k^*)<R_k^{(\min)},
\cr \dfrac{t_{k,n}}{\sum_{n\in \bar{A}_k} t_{k,n}} R_k^{(\max)}, & \sum_{n\in \bar{A}_k} \widetilde{h}(\xi_k^*)>R_k^{(\max)},
\end{cases}
\end{equation}
for $\forall n\in \bar{A}_k$, where  
\begin{equation}
\widetilde{h}(\xi_k^*)=\begin{cases}
 \dfrac{B t_{k,n}}{\ln 2} \log(u_k \xi_k^*) , \!&u_k\ge 1, \cr 0, & u_k<1, 
\end{cases}
\end{equation}
 $u_k=\dfrac{3\gamma C_k^3 g_k B }{ T_k^2 N_0 \ln 2}$, and $=\xi_k^*>0$ is the solution to $\widetilde{U}_k(\xi_k)=0$ with  
\begin{equation*}
\widetilde{U}_k(\xi_k)\overset{\triangle}{=}\l(L_k-\sum_{n\in \bar{A}_k} \widetilde{h}(\xi_k)\r)^2-\xi_k.
\end{equation*}}
\end{proposition}
This proposition shows that given the offloading time division, if $R_k^{(\min)}\le \sum_{n\in \bar{A}_k} \widetilde{h}(\xi_k^*)\le R_k^{\max}$, the optimal offloading policy for the offloading data size has a \emph{threshold-based} structure. Specifically, the mobile offloads partial input data or performs full local computing if $u_k$ is above or below the threshold $1$, respectively. This is expected since offloading can reduce energy consumption only under the conditions of a good channel, stringent latency requirement  or high local-computing complexity. 
\subsubsection{Energy-Efficient Time Division}
Similar to Section~\ref{Sec:GeneTimeDiv}, for any given offloading data sizes $\{\ell_{k,n}\}$, Problem P4 reduces to the following $N$ parallel sub-problems:
\begin{equation}\tag{$\textbf{P6}$}
\begin{aligned}
\min_ {\{t_{k,n}\ge0\} }   &~ \sum_{k\in \bar{B}_n} \frac{t_{k,n}}{g_k}\psi\!\l(\frac{\ell_{k,n}}{t_{k,n}}\r)
\qquad \text{s.t.}
& \sum_{k\in \bar{B}_n} t_{k,n} \le \tau_n, ~~\forall n.
\end{aligned}
\end{equation}
It can be proved that Problem P6 is a convex optimization problem. Define a function $\bar{\psi}(x)$ as  $\bar{\psi}(x)=\psi(x)-x \dfrac{\partial \psi(x)}{\partial x}$. Following the similar procedure as for deriving Proposition~\ref{Pro:GeneTimeDiv}, the optimal time-division policy for this case is characterized as below.
\begin{proposition}\label{Prop:ExpoTime}\emph{Consider asynchronous MECO based on the exponential offloading energy-consumption model. For each epoch, say epoch $n$, given any offloading data sizes $\{\ell_{k,n}\}$, the optimal offloading time division for solving Problem P6, denoted by $\{t_{k,n}^*\}$, is given by
\begin{equation}
t_{k,n}^*=\frac{\ell_k}{\bar{\psi}^{-1}\l( -g_k \eta_k^*\r)},\qquad \forall k\in \bar{B}_n,
\end{equation}
where  $\bar{\psi}^{-1}(x)$ is the inverse function of $\bar{\psi}(x)$  given by $\bar{\psi}^{-1}(x)=\l(B\l(W_0(\frac{x+N_0}{-N_0 e})+1\r)\r)/\ln 2$, and $\eta_k^*>0$ satisfies $\sum_{k\in \bar{B}_n} t_{k,n}^*=\tau_n$.}
\end{proposition}
Last, combining the results of the optimal data partitioning and time division, the optimal solution to Problem P4 can be obtained by an iterative algorithm using the BCD method, which is similar to Algorithm~\ref{Alg:GeneBCD} and omitted for brevity.}}
  
{\color{black}{\subsection{Discussions}
Extension of the proposed BCD solution approach to other more complicated scenarios are discussed as follows.
\begin{itemize}
\item[1)]{\emph{Robust design}:
To cope with imperfect mobile prediction  and  estimation in practice, the current framework can be modified as follows by applying robust optimization techniques. Based on a model of bounded  uncertainty (see e.g., \cite{sharma2015cognitive}), the system-state parameters,  including channel gain, data-arrival time and deadline, can be added with unknown bounded random variables representing estimation-or-prediction  errors. Then using the worst-case approach \cite{sharma2015cognitive},  Problem P1 can be modified by replacing these parameters with their ``worse cases" and then solved using the same  approach, giving a robust offloading policy.
 }
\item[2)]\emph{Online design}: Similar to the online design approach in \cite{chen2007energy}, upon new input-data arrivals or variations of mobiles' information, the proposed control policies can be adjusted by updating information (e.g., new channel gains) and applying the current offline framework to determine the updated data-partitioning and time-division policies. Note that reusing the former results as the initial policy in the iterative recalculation is expected to reduce the computation complexity in temporally-correlated channels. Moreover, the disruptions of task computing can be avoided by continuing the former policy until obtaining the updated one. Last, assuming instantaneous mobiles' information available at the BS, the policy-update approach can also be used for designing the greedy online policy. For frequent arrivals, the computation complexity can be reduced by designing a \emph{random} policy-update approach, where the update probability depends on instantaneous mobiles' information.
\item[3)]{\emph{Time-varying channels}: Assuming block-fading channels where the channel gain is fixed in each fading block and \emph{independent and identically distributed} (i.i.d.) over different blocks, the solution approach can be easily modified that essentially involves re-defining the epoch-set as the fading-block indexes within  the computation duration and the corresponding user-set in Definition~\ref{Def:EpockUser}. Then Problem P1 can be extended by replacing $g_k$ in the objective function with $g_{k,n}$ that denotes the channel gain of mobile $k$ in epoch $n$. This problem can be solved using the same solution approach developed  in the paper.}
\item[4)]{\emph{Non-negligible cloud-computation time}: In the case of non-negligible cloud-computation time, the current problem in Problem P1 can be modified to include the said time in the deadline constraint  as a function of the number of offloaded bits. For example, following the model in \cite{you2016energy}, the cloud-computation time is a linear function of the number of offloaded bits scaled by the fixed cloud-computation duration per bit. Though this entails more complex problems, the general solution approaches developed in this paper for asynchronous MECO  should be still largely applicable albeit  with possible modifications by leveraging results from existing work that considers cloud-computation time (see e.g., \cite{you2016energy}). 
}
\item[5)]{\emph{OFDMA MECO}: Consider the asynchronous MECO system based on OFDMA. Similar to \cite{you2016energy}, the corresponding energy-efficient resource management can be formulated as a mix-integer optimization problem where the integer constrains arise from sub-channel assignments. Though the optimal solution is intractable, following a standard approach, sub-optimal algorithms can be developed  by relaxing the integer constraints and then rounding the results to give sub-channel assignments.}
\item[6)]{{\color{black}{\emph{Binary offloading}: The current results can be used to design the asynchronous MECO based on binary offloading. Note that the corresponding problem is a mixed-integer optimization problem, which is difficult to solve. To address this issue,  a greedy and low-complexity algorithm can be designed by using \emph{probabilistic offloading}. Particularly, with the obtained results for partial offloading in this work, the offloading probability for each mobile can be set as the ratio between offloaded and total data sizes. Then a set of resource-management samples can be generated, each randomly selecting individual mobiles for offloading following the obtained probability. Last, the sample yielding the minimum total mobile energy consumption gives the greedy policy. It is worthy mentioning that the policy can be further improved by using the \emph{cross-entropy} method, which  adjusts the offloading probability based on the outcomes of samples, but  it  will result in higher computation complexity \cite{zhang2014dynamic}.}}}
\end{itemize}  
  }}
  \vspace{-5pt}
\section{Optimal Resource Management with Identical Arrival-Deadline Orders}\label{Sys:Iden}
To gain further insights for the {\color{black}{structure of the optimal resource-management policy}}, this section considers the special case of asynchronous MECO with identical arrival-deadline orders, i.e., a mobile with earlier data arrival also needs to complete the computation earlier. {\color{black}{This case arises when the mobiles have similar computation tasks (e.g., identical online gaming applications) but with random arrivals.}} For this case, the solution to Problem P1 can be further simplified by firstly determining an optimal scheduling order and then designing energy-efficient joint data-partitioning and  time-division policy given  the optimal order. {\color{black}{Note that this design approach does not require the resource management in each epoch.}} We consider that the mobiles and VMs have unbounded computation capacities and the monomial order $m=3$, since it can fairly approximate the transmission-energy consumption in practice.\footnote{The results can be extended to derive the suboptimal policy for the case of $m\neq 3$ by using approximating techniques, although the corresponding optimal policy has no closed form which can be computed by iterative algorithms.} {\color{black}{More importantly, it will lead to useful insights into {\color{black}{the structure of the optimal policy}} as shown in the sequel that the optimal time-division policy admits a defined effective computing-power balancing structure}}. Moreover, the optimal policy is simplified for a  two-user case.

First, we define the \emph{offloading scheduling order} as follows.
\begin{definition}[Offloading Scheduling Order]\label{Def:Order}\emph{Let $\boldsymbol{\theta}=\{\theta_1,\theta_2,\cdots, \theta_I\}$ denote the offloading scheduling order with $\theta_i \in \mathcal{K}$ for $i=1, 2, \cdots, I$.
Under this order,  mobile $\theta_1$ is firstly scheduled for offloading, followed by mobile $\theta_2$,  mobile $\theta_3$ until mobile $\theta_I$. Note that $I\geq K$ in general since each mobile can be scheduled more than once.}
\end{definition} 
\begin{figure}[t]
\begin{center}
\includegraphics[height=3.2cm]{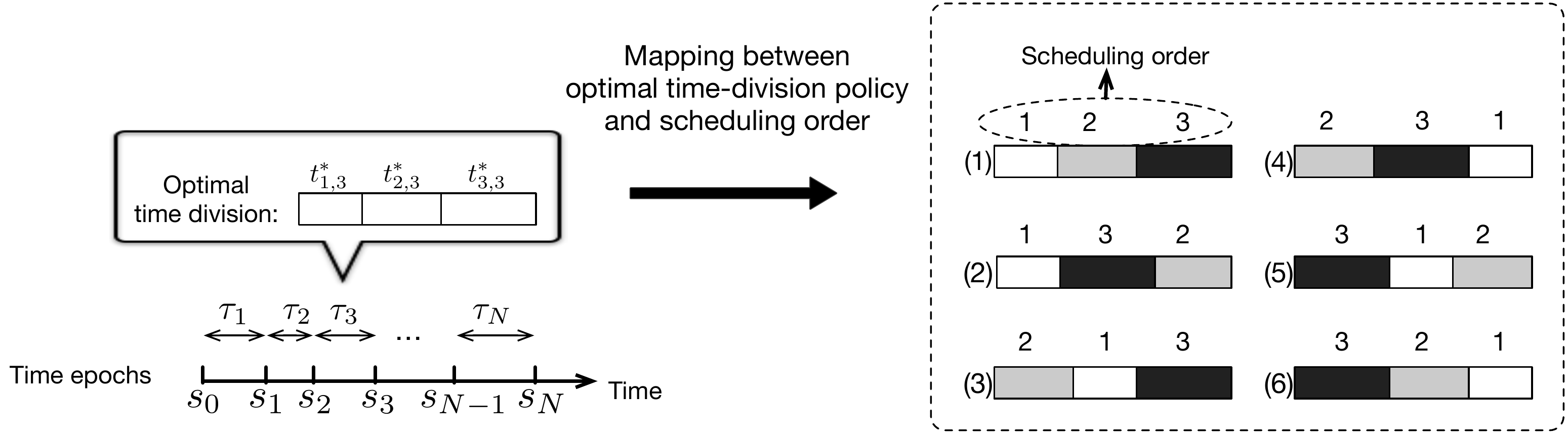}
\caption{Mapping between the time-division policy and scheduling order.}
\label{Fig:TimeOrder}
\end{center}
\end{figure}
Note that given a scheduling order (e.g., $\boldsymbol{\theta}=\{1,2,3,1\}$), one specific mobile (e.g., mobile $1$) can be \emph{repeatedly} scheduled, corresponding to computation offloading  in multiple non-overlapping epochs. Recall that Problem P1 optimizes the offloading time divisions $\{t_{k,n}\}$ and offloaded data sizes $\{\ell_{k,n}\}$ for the mobiles in all epochs. {\color{black}{Specifically, for each epoch, the derived time-division policy only determines the offloading durations allocated for different mobiles, without specifying the scheduling order.  In other words, if considering the scheduling order, one time-division policy resulted from the solution to Problem P1 can correspond to multiple scheduling orders as illustrated in Fig.~\ref{Fig:TimeOrder}. On the other hand, if given the scheduling order, the time-division policy for solving Problem P1 can be uniquely determined.}}

Based on the above definition and discussions, in the following subsections, we first derive one optimal scheduling order and then optimize the joint data-partitioning and time-division policy given the optimal order.
\vspace{-10pt}
 \subsection{Optimal Scheduling Order}
 Recall that given the identical arrival-deadline orders, we have $T_1^{(a)}\le T_2^{(a)}\le \cdots \le T_K^{(a)}$ and $T_1^{(d)}\le T_2^{(d)}\le \cdots \le T_K^{(d)}$. This means that mobile $(k-1)$ has \emph{earlier} data arrival than mobile $k$ and also requires the computation to be completed earlier.  Using this key fact, we characterize one optimal offloading scheduling order as follows, which is proved in Appendix~\ref{App:OptOrder}.
 
\begin{lemma}[Optimal Scheduling Order]\label{Lem:OptOrder}\emph{For the case of identical arrival-deadline order, one optimal scheduling order that can lead to the optimal solution to Problem P1 is $\boldsymbol{\theta}^*=\{1,2, \cdots K\}$.
}
\end{lemma}
{\color{black}{Lemma~\ref{Lem:OptOrder} shows that for the case of identical arrival-deadline orders, there exists one optimal deterministic and simple scheduling order that entails \emph{sequential transmission}  by mobiles following their data-arrival order. The intuitive reason behind the optimality of such an order is that the mobile with an earlier  input-data arrival  has a more pressing  deadline. On the other hand,  for the case with general arrival-deadline orders, the optimal scheduling  has no clear  structure, due to the irregularity in data arrivals and deadlines across  mobiles.
 }}
\vspace{-8pt}
\subsection{Energy-Efficient Data Partitioning and Time Division Given the Optimal Scheduling Order}
Given the optimal  scheduling order in Lemma~\ref{Lem:OptOrder}, this subsection aims to jointly optimize the offloaded data sizes $\{\ell_{k}\}$ and offloading durations $\{t_k\}$ for the mobiles for achieving the minimum total mobile-energy consumption.  

{\color{black}{Note that, instead of partitioning each epoch duration for relevant mobiles, the introduced scheduling order helps provide  an alternative design approach that can directly partition the total time interval $\l[0, T_K^{(d)}\r]$  for the mobiles given the optimal scheduling order. This approach yields new insights for the policy structure as elaborated in the sequel.}} Specifically, let $t_k^{(s)}$, $t_k$ and $\ell_k$ denote the starting-time instant, total offloading duration and offloaded data size for mobile $k$, respectively. The offloading for the mobiles should satisfy the following constraints. First, under the \emph{data causality constraint} which prohibits input data from being offloaded before it arrives, we have 
\begin{equation}
\vspace{-5pt}
\text{(Data causality constraint)}\quad t_k^{(s)}\ge T_k^{(a)}, \quad\forall k.
\end{equation} Next, the deadline constraint requires that
\begin{equation}
\vspace{-5pt}
 \text{(Deadline constraint)}\quad t_k^{(s)}+ t_{k}\le T_k^{(d)}, \quad\forall k.
 \end{equation} 
 In addition, the time-sharing constraint in \eqref{Eq:TimeShare} reduces to the \emph{time non-overlapping constraint} as:
 \begin{equation}
 \vspace{-5pt}
 \text{(Time non-overlapping constraint)}\quad t_{k+1}^{(s)}\ge t_k^{(s)}+ t_{k}, \quad\forall k,
 \end{equation} 
 where $t_{K+1}^{(s)}$ is defined as $t_{K+1}^{(s)}=T_K^{(d)}$.
Based on Lemma~\ref{Lem:OptOrder} and above constraints, the solution to Problem P1 assuming $m=3$ can be derived by solving the following problem:
\begin{equation}\tag{$\textbf{P4}$} 
\begin{aligned}
\min_ {\l\{t_k^{(s)} \ge 0, \ell_{k}\ge0, t_{k}\ge0\r\} }   ~ &\sum_{k=1}^{K} \l[ \frac{\lambda(\ell_{k})^3}{g_k(t_{k}) ^{2}} + \dfrac{\gamma C_k^3(L_k-\ell_{k})^3}{T_k^2} \r] \\
\text{s.t.}\qquad 
&  \ell_{k}\le L_k, ~~~  t_k^{(s)}\ge T_k^{(a)}, ~~~t_k^{(s)}+ t_{k}\le \min\l\{T_k^{(d)}, t_{k+1}^{(s)}\r\},  & \forall k.
\end{aligned}
\end{equation} 
Problem P4 can be proved to be a convex optimization problem using the similar method as for deriving Lemma~\ref{Lem:P1Conv}. One important property of Problem P4 is given below, which can be proved by contradiction and the proof is omitted for brevity.
\begin{lemma}\label{Lem:UseAllTime}\emph{For the case of identical arrival-deadline orders, the optimal offloading starting-time instants and durations for solving Problem P4, denoted by $\l\{\l(t_k^{(s)}\r)^*, t_k^*\r\}$, satisfy the following:
\begin{equation}
\l(t_k^{(s)}\r)^*=\begin{cases}
 \sum_{i=1}^{k-1} t_i^*, & k>1, \cr 0, &k=1, 
\end{cases}
 \qquad \text{and}\qquad \sum_{k=1}^K t_k^*=T_K^{(d)}.
\end{equation} }
\end{lemma}
Lemma~\ref{Lem:UseAllTime} indicates that the multiuser offloading should \emph{fully} utilize the whole time duration, which is expected since offloading-energy consumption decreases with the offloading duration. Using Lemma~\ref{Lem:UseAllTime}, Problem P4 can be rewritten as follows.
\begin{equation}\tag{$\textbf{P5}$} 
\vspace{-5pt}
\begin{aligned}
\min_ {\{\ell_{k}\ge0, t_{k}\ge0\} }   ~ &\sum_{k=1}^{K}  \l[ \frac{\lambda(\ell_{k})^3}{g_k(t_{k}) ^{2}} + \dfrac{\gamma C_k^3(L_k-\ell_{k})^3}{T_k^2} \r] \\
\text{s.t.}\quad 
& \ell_{k}\le L_k,  \qquad&&\forall k,\\
&T_{k+1}^{(a)}\le \sum_{i=1}^k t_i \le T_k^{(d)}, && k=1, 2, \cdots, K-1,\\
&\sum_{k=1}^K t_k=T_K^{(d)}.
\end{aligned}
\end{equation} 
Note that given the constraint of $\l(T_{k+1}^{(a)}\le \sum_{i=1}^k t_i\r)$, the data causality constraint is always satisfied since $t_{k+1}^{(s)}=\l(\sum_{i=1}^{k-1} t_i\r)+ t_{k}\ge T_{k+1}^{(a)}$. Moreover, $\sum_{i=1}^k t_i \le T_k^{(d)}$ indicates the deadline constraint. It can be easily proved that Problem P5 is a convex optimization problem. To characterize {\color{black}{the structure of the optimal policy}}, we decompose Problem P5 into two sub-problems, namely the slave problem corresponding to the energy-efficient data partitioning given offloading durations and the master one for the energy-efficient time division. 
\subsubsection{Slave Problem for Energy-Efficient Data Partitioning Given Offloading Durations}\label{Sec:SingleData} For any given offloading durations $\{t_{k}\}$, Problem P5 reduces to the slave problem that optimizes the offloaded data sizes $\{\ell_k\}$. It is easy to see that this slave problem can be decomposed into $K$ parallel subproblems as
 \begin{equation}\tag{$\textbf{P6}$} 
\begin{aligned}
\min_{\ell_k} \quad&  \frac{\lambda(\ell_{k})^3}{g_k(t_{k}) ^{2}} + \dfrac{\gamma C_k^3(L_k-\ell_{k})^3}{T_k^2} 
\qquad\text{s.t.}\quad 
& 0\le \ell_k \le L_k.
\end{aligned}
\end{equation}
 Problem P6 is a convex optimization problem and the optimal solution can be derived in the following proposition by using  the Lagrange method.
\begin{proposition}\label{Prop:OptDataPart}\emph{For each mobile, say mobile $k$, given the offloading duration $t_k$, the energy-efficient data partitioning policy
  is given by $\ell_k^*=\dfrac{\theta_k}{1+\theta_k} L_k$,
where $\theta_k=\sqrt{\dfrac{b_k}{a_k}}\dfrac{t_k}{T_k}$. The corresponding minimum mobile-energy consumption, denoted by $E_k^*$, is a function of $t_k$:
\begin{equation}\label{Eq:EachOptEgy}
E_k^*(t_k)=\dfrac{a_kL_k^3}{\l(\sqrt{\dfrac{a_k}{b_k}}T_k+t_k\r)^2}.
\end{equation}}
\end{proposition}
{\color{black}{Note that the minimum mobile-energy consumption given offloading duration, $E_k^*(t_k)$, has a simple form. This can facilitate solving the master problem for time division  in the sequel.}}
\begin{remark}[Proportional Offloading]\emph{Proposition~\ref{Prop:OptDataPart} means that the offloaded data size is \emph{proportional} to the total data size $L_k$ with a proportional factor $\dfrac{\theta_k}{1+\theta_k}$, which is jointly determined by the channel gain $g_k$, offloading time duration $t_k$ and computation deadline $T_k$. It can be inferred 
that more data should be offloaded for a longer offloading duration or better channel. }
\end{remark}
\subsubsection{Master Problem for Energy-Efficient Time Division}\label{Sec:MulRA} Using the result of energy-efficient data partitioning in the preceding subsection, the master problem focuses on multiuser time division for minimizing the total mobile-energy consumption. Using Proposition~\ref{Prop:OptDataPart}, Problem P5 can be equivalently reduced to the following problem.{\footnote{{\color{black}{After problem transformations, the resulted   Problem P7 focusing on offloading time division optimization has a similar form with the optimization problems  in \cite{zafer2005calculus,chen2007energy} for asynchronous data transmissions, which reveals the intrinsic connections between these two areas.}}}}.
\vspace{-8pt}
 \begin{equation}\tag{$\textbf{P7}$} 
\begin{aligned}
\min_{\{t_k\ge 0\}} \quad&\sum_{k=1}^K \frac{a_kL_k^3}{\l(\sqrt{\dfrac{a_k}{b_k}}T_k+t_k\r)^2}\\
\text{s.t.}\quad 
&T_{k+1}^{(a)}\le \sum_{i=1}^k t_i \le T_k^{(d)}, \quad k=1, 2, \cdots, K-1,\\
&\sum_{k=1}^K t_k=T_K^{(d)}.
\end{aligned}
\end{equation}

Problem P7 is a convex optimization problem and the corresponding Lagrangian is 
\begin{equation}
L=\sum_{k=1}^K \frac{a_k L_k^3}{\l(\sqrt{\dfrac{a_k}{b_k}}T_k+t_k\r)^2}+\sum_{k=1}^{K-1} \mu_k\l(T_{k+1}^{(a)}-\sum_{i=1}^k t_i\r)+\sum_{k=1}^K \omega_k \l(\sum_{i=1}^k t_i-T_k^{(d)}\r)-\sum_{k=1}^K \sigma_k  t_k,
\end{equation} 
where $(\mu_k\geq 0)$, $(\omega_k\geq 0)$, and $(\sigma_k\ge 0)$ correspond to the Lagrange multipliers for the constraints of $\l(T_{k+1}^{(a)}-\sum_{i=1}^k t_i\le 0 \r)$, $\l(\sum_{i=1}^k t_i-T_k^{(d)}\le 0\r)$, and $\l(t_k\ge0\r)$, respectively.
For ease of notion, define a \emph{reference} function $f_k(a_k, b_k, x)$ as: 
\begin{equation}\label{Eq:RefFunc}
f_k(a_k, b_k, x)=(a_kL_k^3)\bigg/\l(\sqrt{\dfrac{a_k}{b_k}}T_k+x\r)^3.
\end{equation} Then applying the \emph{Karush-Kuhn-Tucker} (KKT) conditions leads to the following sufficient and  necessary conditions for the optimality of Problem P7:
\begin{subequations}
\begin{align}
&\dfrac{\partial L}{\partial t_k^*}=
-2f_k(t_k^*)-\sum_{i=k}^{K-1} \mu_i^* +\sum_{i=k}^K \omega_i^*-\sigma_k^*=0,  \qquad\qquad \quad \forall k;\label{Eq:KKT1}\\
& \mu_k^*\l(T_{k+1}^{(a)}-\sum_{i=1}^k t_i^*\r)=0,   ~~~\quad\quad \qquad \qquad \qquad\qquad \qquad k=1, 2, \cdots, K-1; \label{Eq:KKT2}\\
&\omega_k^* \l(\sum_{i=1}^k t_i^*-T_k^{(d)}\r)=0,~~\sigma_k^* t_k^*=0, ~~\quad\quad \qquad\quad\qquad\quad \forall k; \label{Eq:KKT3}\\
&\sum_{k=1}^K t_k^*=T_k^{(d)}, \label{Eq:KKT4}
\end{align}
\end{subequations}
where $\sum_{i=K}^{K-1}\mu_i^*\overset{\triangle}{=}0$.
 Combing the conditions in \eqref{Eq:KKT1}-\eqref{Eq:KKT4} yields the following key results.
\begin{proposition}\label{Prop:OptMulRes}\emph{Consider the case of identical arrival-deadline orders. Given the optimal scheduling order  in Lemma~\ref{Lem:OptOrder}, the optimal time-division policy can be expressed as:
\vspace{-7pt}
\begin{equation}
t_k^*=\l(\frac{2a_k L_k^3}{\sum_{i=k}^K \omega_i^*-\sum_{i=k}^{K-1}\mu_i^*-\sigma_k^*}\r)^{\frac{1}{3}}-\sqrt{\frac{a_k}{b_k}}T_k, \quad\forall k,
\end{equation}
where $\l\{\mu_k^*,\omega_k^*, \sigma_k^*\r\}$ satisfy the conditions in \eqref{Eq:KKT2}-\eqref{Eq:KKT4}.}
\end{proposition}
Before characterizing {\color{black}{the structure of the optimal policy}}, we first introduce several effective computing parameters in the following.
\begin{definition}[Effective Computing Parameters]\emph{Let $T_k^{(\rm{eff})}$ denote the \emph{effective computing duration} for mobile $k$, defined as the weighted sum of the local-computing and offloading durations, given by $T_k^{(\rm{eff})}\overset{\triangle}{=}\sqrt{\dfrac{a_k}{b_k}}T_k+t_k^*$. In addition, let $P_k^{(\rm{eff})}$ denote the \emph{effective computing power} defined by
\vspace{-7pt}
\begin{equation}\label{Eq:EffectPower}
\text{(Effective computing power)}\quad P_k^{(\rm{eff})}\overset{\triangle}{=}\frac{E_k^*}{T_k^{(\rm{eff})}}.
\end{equation}}
\end{definition} 
 The defined effective computing duration can be intuitively interpreted  as the allocated computing duration in a \emph{combined} CPU with parallel local-computing and offloading components. The weighting factor $\sqrt{\dfrac{a_k}{b_k}}$ represents the effective offloading duration for $1$-second local-computing duration.  Moreover,  substituting \eqref{Eq:EachOptEgy} into \eqref{Eq:EffectPower} and then combing it with \eqref{Eq:RefFunc} yields that $P_k^{(\rm{eff})}=f_k(a_k, b_k, t_k^*)$.  Using these definitions, \eqref{Eq:KKT1} can be rewritten as 
\begin{equation}\label{Eq:EffComPower}
P_k^{(\rm{eff})}=\frac{\sum_{i=k}^K \omega_i^*-\sum_{i=k}^{K-1}\mu_i^* -\sigma_k^*}{2}.
\end{equation}
Then the important properties of $\l\{P_k^{(\rm{eff})}\r\}$ are characterized in the following corollary, which can be directly proved by considering the conditions in \eqref{Eq:KKT2}-\eqref{Eq:KKT3} and according to \eqref{Eq:EffComPower}. 
\begin{corollary}[Properties of Effective Computing Power]\label{Cor:PowProp}\emph{Consider the case of identical arrival-deadline orders. Given the optimal time-division policy in Proposition~\ref{Prop:OptMulRes} under the optimal scheduling order given in Lemma~\ref{Lem:OptOrder}, let $\mathcal{\bar{K}}$ denote the mobile indexes allocated with offloading durations, given by $\mathcal{\bar{K}}=\{k~|~t_k^*>0\}$. The corresponding effective computing power expressed in \eqref{Eq:EffComPower}, has the following structures:
\begin{itemize}
\item[1)] If all mobiles have identical data-arrival time instants, i.e., $T_1^{(a)}=T_2^{(a)}=\cdots=T_K^{(a)}$, for the mobiles allocated with offloading durations, the effective computing power is \emph{monotonically-decreasing}, i.e., $P_k\ge P_{j}$, for $k<j$ and $\{k,j\}\subseteq \mathcal{\bar{K}}$.
\item[2)] If all mobiles have identical computation deadlines, i.e., $T_1^{(d)}=T_2^{(d)}=\cdots=T_K^{(d)}$, for the mobiles allocated with offloading durations, the effective computing power is \emph{monotonically-increasing}, i.e., $P_k\le P_{j}$, for $k<j$ and $\{k,j\}\subseteq \mathcal{\bar{K}}$.
\item[3)] Consider two consecutive mobiles allocated with offloading durations, i.e., $\{k,k+1\}\subseteq \mathcal{\bar{K}}$. For mobile $k$, if it satisfies $T_{k+1}^{(a)}< \sum_{i=1}^k t_i^* < T_k^{(d)}$, its effective computing power is the same as that of the subsequently-scheduled mobile, i.e., $P_k^{(\rm{eff})}=P_{k+1}^{(\rm{eff})}$. Otherwise, $P_k^{(\rm{eff})}\le P_{k+1}^{(\rm{eff})}$ if $\sum_{i=1}^k  t_i^*=T_{k+1}^{(a)}$, and  $P_k^{(\rm{eff})}\ge P_{k+1}^{(\rm{eff})}$ if $\sum_{i=1}^k  t_i^*=T_{k}^{(d)}$.
\end{itemize}}
\end{corollary}
In Corollary~\ref{Cor:PowProp}, the monotonicity in case 1) shows that given the same data-arrival instants, for the offloading mobiles, the later the deadline, the smaller the effective computing power. In particular, if the deadline constraint  for mobile $k$ is inactive (i.e., $\sum_{i=1}^k t_i^*<T_k^{(d)}$), it indicates that this mobile has relatively loose deadline requirement. To achieve the minimum total mobile-energy consumption, instead of simply reducing its own energy consumption, it is more energy-efficient for mobile $k$ to spare partial time duration for reducing energy consumption of the later mobile $j$. By doing so, these two mobiles share the same effective computing power. Otherwise, mobile $k$ consumes larger effective computing power than mobile $j$, since it has quite stringent deadline constraint. The structure in case 2) reflects a similar principle that given the same deadline,  the mobile with earlier-arrived input data tends to consume smaller effective computing power.  Case 3) considers the general case of different data-arrival time instants and deadlines. It can be observed that for the mobiles allocated with offloading durations, only when both the data causality constraint for mobile $(k+1)$ and deadline constraint for mobile $k$ are inactive,  mobile $k$ shares the same effective computing power with mobile $(k+1)$. Otherwise, mobile $k$ consumes smaller and larger effective computing power than mobile $(k+1)$, if the data causality constraint for mobile $(k+1)$ and deadline for mobile 
$k$ is active, respectively.
\vspace{-5pt}
\begin{remark}[Computing-Power Balancing]\label{Rem:ComPowBalan}\emph{Corollary~\ref{Cor:PowProp} indicates that the optimal time-division policy tends to \emph{balance the effective computing power} among \emph{offloading mobiles} via time sharing. The variations of effective computing power arise from  the activeness of the data causality and deadline constraints. In particular, if the mobiles have identical arrival time instants and deadlines, the identical effective computing-power policy for offloading mobiles is the optimal policy.}
\end{remark}
\vspace{-12pt}
\subsection{Two-User Case}
This subsection studies the offloading policy for a special case with $K=2$, referred to as the  two-user scheduling.  Without loss of generality for the case of identical arrival-deadline orders, we assume that $0=T_1^{(a)}<T_2^{(a)}<T_1^{(d)}<T_2^{(d)}$ (see Fig.~\ref{Fig:System}). These two mobiles time-share one \emph{common} time interval, $\l[T_2^{(a)}, T_1^{(d)}\r]$, for computation offloading. Given the optimal scheduling order in Lemma~\ref{Lem:OptOrder} and data-partitioning policy in Proposition~\ref{Prop:OptDataPart}, the problem for the energy-efficient two-user time division  can be formulated as below by simplifying Problem P7.
 \begin{equation}\tag{$\textbf{P8}$} 
\begin{aligned}
\min_{\{t_1,t_2\}} \quad& \frac{a_1L_1^3}{\l(\sqrt{\dfrac{a_1}{b_1}}T_1+t_1\r)^2}+\frac{a_2L_2^3}{\l(\sqrt{\dfrac{a_2}{b_2}}T_2+t_2\r)^2}\\
\text{s.t.}\quad 
& t_k^{(\min)}\le t_k \le t_k^{(\max)}, \quad k=1,2,\\
& t_1+t_2=T_2^{(d)},\\
\end{aligned}
\end{equation}
where $t_1^{(\min)}=T_{2}^{(a)}$, $t_1^{(\max)}=T_1^{(d)}$, $t_2^{(\min)}=T_{2}^{(d)}-T_1^{(d)}$, and $t_2^{(\max)}=T_2^{(d)}-T_2^{(a)}$.

To characterize {\color{black}{the structure of the optimal policy}}, we first  give the  properties of the function $f_k(a_k, b_k, x)$  in the following, which can  be easily proved, thus the proof is omitted for brevity.
\begin{lemma}\label{Lem:PropF}\emph{The function $f_k(a_k, b_k, x)$ has the following properties:
\begin{itemize}\itemsep=-3pt
\item[1)] $f_k(a_k, b_k, x)$ is  monotonically-decreasing with $x$ and monotonically-increasing with $b_k$.
\item[2)] $f_k(a_k, b_k, x)$ is monotonically-increasing with $a_k$ for $a_k\le 4b_k x^2/T_k^2$ and monotonically-decreasing for $a_k>4b_k x^2/T_k^2$.
\end{itemize} }
\end{lemma}
Then, we define the \emph{offloading  region} for each mobile denoted by $\mathcal{G}_k=[d_{k1}, d_{k2}]$, where 
\begin{equation*}
\vspace{-5pt}
d_{k1}=f_k\l(a_k, b_k, t_k^{(\max)}\r),~~\text{and}~~d_{k2}=f_k\l(a_k, b_k, t_k^{(\min)}\r).
\end{equation*} 
Using Lemma~\ref{Lem:PropF},  we have $d_{k1}<d_{k2}$. Thus, $d_{k1}$ and $d_{k2}$ can be interpreted as the minimum and maximum achievable effective computing power for mobile $k$, respectively.

 Based on the above definitions, the energy-efficient time-division policy  is given as follows.
\begin{corollary}[Optimal Two-User Time Division]\label{Cor:TwoUser}\emph{For the two-user case, given the optimal scheduling order $\boldsymbol{\theta}^*=\{1,2\}$, the energy-efficient time-division policy  is given by
\begin{itemize}
\item[1)] If $d_{11}\ge d_{22}$, we have $t_1^*=t_1^{(\max)}$ and $t_2^*=t_2^{(\min)}$.
\item[2)] If $d_{12}\le d_{21}$,  we have $t_1^*=t_1^{(\min)}$ and $t_2^*=t_2^{(\max)}$.
\item[3)] Otherwise, 
\vspace{-7pt}
\begin{equation}
t_k^*=\l(\frac{2a_k L_k^3}{\omega^*}\r)^{\frac{1}{3}}-\sqrt{\frac{a_k}{b_k}}T_k, \quad  k=1, 2,
\end{equation}
where $\omega^*$ satisfies $t_1^*+t_2^*=T_2^{(d)}$.
\end{itemize}}
\end{corollary} 
Corollary~\ref{Cor:TwoUser} can be easily proved by the Lagrange method. It reveals that  the optimal two-user time-division policy has a \emph{double-threshold} structure. Specifically, mobile $1$ \emph{fully occupies} the common time interval if its minimum effective computing power is larger than the maximum effective computing power of mobile $2$ (i.e., $d_{11}\ge d_{22}$), and \emph{does not share} the common interval if its maximum effective computing power is smaller than the minimum effective computing power of mobile $2$  (i.e., $d_{12}\le d_{21})$. Otherwise, both the mobiles time-share the common duration, achieving the same effective computing power.

\begin{remark}[Effects of Parameters on Two-User Scheduling]\emph{Combing Lemma~\ref{Lem:PropF} and Corollary~\ref{Cor:TwoUser} and using the definition of $\{a_k, b_k\}$ in Lemma~\ref{Prop:OptDataPart}, we can observe that the mobile with a higher computation complexity  (i.e., larger $C_k$) tends to be allocated with a longer offloading duration, since it requires more CPU cycles. On the other hand, as the channel gain $g_k$ grows, the allocated offloading duration is firstly increasing and then deceasing after exceeding a threshold. This observation can be interpreted as follows. If the channel is relatively poor, increasing the channel gain can significantly reduce the transmission-energy consumption and thus a longer offloading duration is preferred. However, when the channel gain exceeds a certain threshold, increasing its offloading duration can no longer substantially achieve energy savings, such that it is better to spare a longer duration for the other mobile to reduce the total mobile-energy consumption. }
\end{remark}

\vspace{-10pt}
\section{Optimal Resource Management with Reverse Arrival-Deadline Orders} \label{Sys:Rever}

{\color{black}{In this section, we consider another special case with reverse arrival-deadline orders, i.e., a mobile with input data arriving later needs to complete the computation earlier.  This  may model the practical scenario on mixing mobiles with latency-tolerant applications and those with latency-critical applications.}}  Specifically, we derive the optimal scheduling order and propose a transformation-and-scheduling approach to derive the optimal offloading control.

\vspace{-5pt}
\subsection{Optimal Scheduling Order}
Recall that in this case, the data-arrival time instants and deadlines for different mobiles follow the orders of: $T_1^{(a)}\le T_2^{(a)}\le \cdots \le T_K^{(a)}$ and $T_1^{(d)}\ge T_2^{(d)}\ge \cdots \ge T_K^{(d)}$. 
 To solve Problem P1, we first present one optimal scheduling order for this case, given in the following lemma.
\begin{lemma}\label{Lem:OptOrderRev}\emph{For the case of reverse arrival-deadline orders, one optimal scheduling order that can lead to the optimal solution to Problem P1 is $\boldsymbol{\theta}^*=\{1,2, \cdots, K-1, K, K-1, \cdots, 2,1\}.$}
\end{lemma}
Lemma~\ref{Lem:OptOrderRev} is proved in Appendix~\ref{App:OptOrderRev}. It can be intuitively interpreted that the optimal scheduling order is composed of two sub-orders in the durations of $\l[0, T_K^{(d)}\r]$ and $\l[T_K^{(d)}, T_1^{(d)}\r]$, corresponding to $\{1,2, \cdots, K\}$ and $\{K-1, \cdots, 2,1\}$, respectively.

\vspace{-5pt}
\subsection{Energy-Efficient Data Partitioning and Time Division Given the Optimal Scheduling Order}
Note that given the optimal scheduling order in Lemma~\ref{Lem:OptOrderRev}, each mobile (except mobile $K$) is scheduled twice. 
This renders the previous approach of direct multiuser time division without considering the order no longer inapplicable for the current case. To address this issue, we propose a \emph{transformation-and-scheduling} approach to derive the optimal data-partitioning and time-division policies as detailed in the sequel.  
\subsubsection{Transformation} This phase aims to transform  the original problem into the counterpart of identical arrival-deadline orders under the condition of \emph{preserving} the time-sharing relationship (referring to the overlapping durations) among mobiles, such that the optimal offloading duration for each mobile can be derived by the developed time-division policy in Section~\ref{Sec:MulRA}. This essentially involves a proposed \emph{deadline-alignment migration} technique defined below.
\begin{definition}[Deadline-Alignment Migration]\label{Def:DAM}\emph{The deadline-alignment migration scheme imposes the same deadline for all the mobiles by migrating the computing time interval of each mobile from $\l[T_k^{(a)}, T_k^{(d)}\r]$ to $\l[T_k^{(a)}+\Delta_k, T_1^{(d)}\r]$, where $\Delta_k=T_1^{(d)}-T_k^{(d)}$.}
\end{definition}
 Under this scheme, Problem P1 can be transformed to the following problem.
\begin{equation}\tag{$\textbf{P9}$} 
\begin{aligned}
\min_{\l\{t_k^{(s)} \ge 0, \ell_{k}\ge0, t_{k}\ge0\r\} }   ~ &\sum_{k=1}^{K} \l[ \frac{\lambda(\ell_{k})^3}{g_k(t_{k}) ^{2}} + \dfrac{\gamma C_k^3(L_k-\ell_{k})^3}{T_k^2} \r] \\
\text{s.t.}\qquad 
&  \ell_{k}\le L_k, ~~~ t_k^{(s)}\ge T_k^{(a)}+\Delta_k, ~~~t_k^{(s)}+ t_{k}\le \min\l\{T_1^{(d)}, t_{k+1}^{(s)}\r\},  & \forall k.
\end{aligned}
\end{equation} 
One can observe that Problem P9 has the same form as Problem P4 and only differs in the values of data-arrival time instants and deadlines. Hence, it can be solved using the same solution approach developed in Section~\ref{Sec:MulRA}, with details omitted for brevity. The corresponding optimal data-partitioning and time-division policy is denoted by $\{t_k^*,\ell_k^*\}$.
\subsubsection{Scheduling} Given the optimal total offloaded data size and offloading duration for each mobile derived in the \rm{transformation} phase, the scheduling phase focuses on allocating the offloading time intervals and offloaded data sizes given the optimal scheduling order in Lemma~\ref{Lem:OptOrderRev}. To this end, we propose a scheduling approach called \emph{reverse-order scheduling}, as presented in Algorithm~\ref{Alg:BackSche}. The key idea is to \emph{sequentially} determine the offloading durations for mobile $K$, mobile $(K-1)$, until mobile $1$, 
which accounts for the optimal scheduling order.

The detailed procedures are elaborated as follows.  Step (1) specifies the offloading interval for mobile $K$, which only has one interval. Next, 
Step (2) determines the two offloading intervals of mobile $(K-1)$ that are before and after the time interval of mobile $K$, denoted by $\l[y_{K-1}^{(s)}, y_{K-1}^{(e)}\r]$ and $\l[z_{K-1}^{(s)}, z_{K-1}^{(e)}\r]$, respectively. In particular, it allocates the longest time duration for $\l[y_{K-1}^{(s)}, y_{K-1}^{(e)}\r]$ with duration given by $\min\l\{t_{K-1}^*, y_{K}^{(e)}-T_{K-1}^{(a)}\r\}$. Note that this guarantees that the scheduling satisfies the data causality constraint. The remaining offloading duration of mobile $(K-1)$ is allocated in $\l[z_{K-1}^{(s)}, z_{K-1}^{(e)}\r]$. Similarly, other mobiles' offloading intervals can be determined following the same procedure. Last, for each user, the offloaded data sizes in the two scheduling intervals are allocated proportionally to the duration length. 
\begin{algorithm}[t]
  \caption{The Proposed Reverse-Order Scheduling.}
  \label{Alg:BackSche}
    \begin{itemize}
  \item\textbf{Step 1} [Initialize]: Let $k=K$ and the offloading time interval is $\l[y_k^{(s)}, y_k^{(e)}\r]$, where $y_k^{(s)}=T_k^{(d)}-t_k^*$, and $y_k^{(e)}=T_k^{(d)}$.
   \item\textbf{Step 2} [Update]: $y_{k-1}^{(e)}=y_k^{(s)}$, $z_{k-1}^{(s)}=y_k^{(e)}$, and $k=k-1$.
  \item\textbf{Step 3} [Reverse-order scheduling]: \emph{While}  ($k>0$)
  \begin{itemize}
  \item [(1)] The offloading time intervals for mobile $k$ are $\l[y_k^{(s)}, y_k^{(e)}\r]$ and $\l[z_k^{(s)}, z_k^{(e)}\r]$, given by: 
  \begin{itemize}
  \item[1)] If $\delta_k\overset{\triangle}{=}y_k^{(e)}-t_k^*\ge T_k^{(a)}$, then $y_k^{(s)}=y_k^{(e)}-t_k^*$ and $z_k^{(e)}=z_k^{(s)}$. 
  \item[2)] Otherwise, $y_k^{(s)}=T_k^{(a)}$ and $z_k^{(e)}=z_k^{(s)}+(T_k^{(a)}-\delta_k)$. 
  \end{itemize}
  \item[(2)] Update: $y_{k-1}^{(e)}=y_k^{(s)}$, $z_{k-1}^{(s)}=z_k^{(e)}$, and $k=k-1$.
  \end{itemize}
  \item\textbf{Step 4}: The offloaded data sizes for mobile $k$ in the intervals of $\l[y_k^{(s)}, y_k^{(e)}\r]$ and $\l[z_k^{(s)}, z_k^{(e)}\r]$, are given by 
   $\ell_k^{*(1)}=\frac{\l(y_k^{(e)}-y_k^{(s)}\r)\ell_k^*}{t_k^*}$ and $\ell_k^{*(2)}=\frac{\l(z_k^{(e)}-z_k^{(s)}\r)\ell_k^*}{t_k^*}$, respectively. 
  \end{itemize}
  \end{algorithm}

{\color{black}{The proposed transformation-and-scheduling approach yields the optimal solution to Problem P1. Essentially, the optimality is due to the fact that the deadline-alignment migration does not change the time-sharing relationship among mobiles and the scheduling phase satisfies both the data-causality and deadline constraints.}}

\vspace{-8pt}
\section{Simulation Results and Discussions}\label{Sys:Simu}
In this section, the performance of proposed resource-management policies for asynchronous MECO systems is evaluated by simulations based on $1000$ realizations. The simulation parameters are set as follows unless specified otherwise. {\color{black}{The MECO system consists of $30$ mobiles}}, which independently generate computation input data in the time interval of $[0, 3]$ s, following the uniform distribution. Moreover, the length of required latency follows the exponential distribution with the expected latency set as $0.6$ s. Both the data size and required number of CPU cycles per bit follow the uniform distribution with $L_k\in[0, 60]$ KB ($10^3$ bits) and $C_k\in [500, 1500]$ cycles/bit. The maximum mobile CPU frequency is uniformly selected from the set $\{0.1, 0.2, \cdots, 1.0\}$ GHz and the maximum VM computation capacity for each mobile is uniformly distributed in $D_k\in [0, 4]\times 10^{9}$ cycles.  The constant $\gamma$ is set as $\gamma=10^{-28}$ \cite{you2015energyJSAC}. For offloading, we set the monomial order $m=3$ and the 
energy coefficient $\lambda=10^{-25}$. The channel power gain $g_k= |h_k|^2$ where $h_k$ is modeled as independent Rayleigh fading with the average power loss set as $10^{-3}$.

{\color{black}{For performance comparison, we consider the following baseline policies. The first one is the \emph{equal time-division} policy that first allocates equal time durations in each epoch for the mobiles that time-share the epoch, and then optimizes the data-partitioning policy for each mobile. The other two are called \emph{one-round} and \emph{two-round iteration} policies that initiate the algorithm with equal time division and then perform one-round and two-round BCD iterations, respectively.
}}
\vspace{-5pt}
\subsection{General Arrival-Deadline Orders}
\begin{figure}[t!]
\centering
\subfigure[Effect of the monomial order.]{\label{Fig:GeneEgy_vs_order}
\includegraphics[width=7cm]{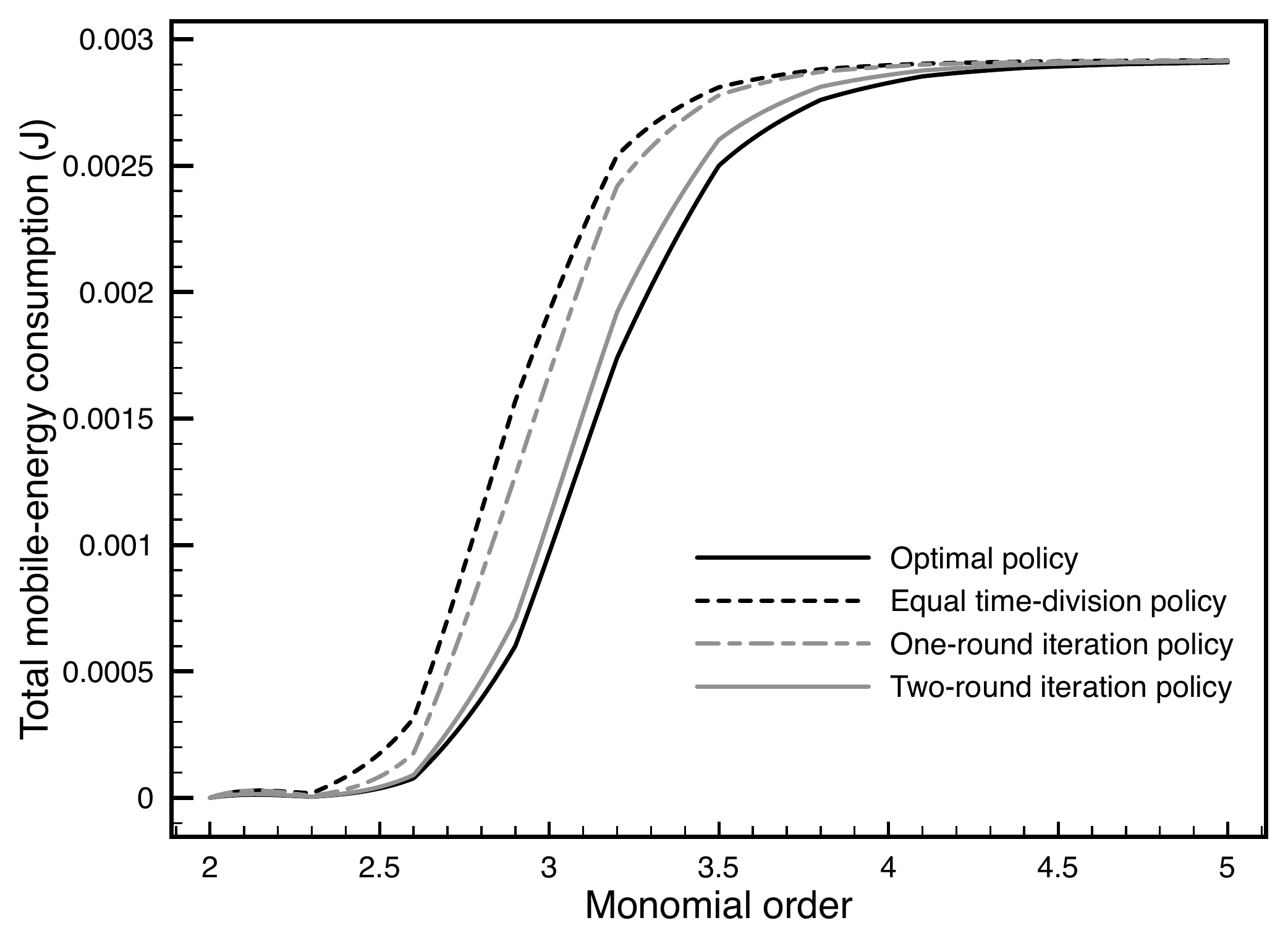}}
\hspace{0.5cm}
\subfigure[Effect of the expected latency requirement.]{\label{Fig:GeneEgy_vs_duration}
\includegraphics[width=7cm]{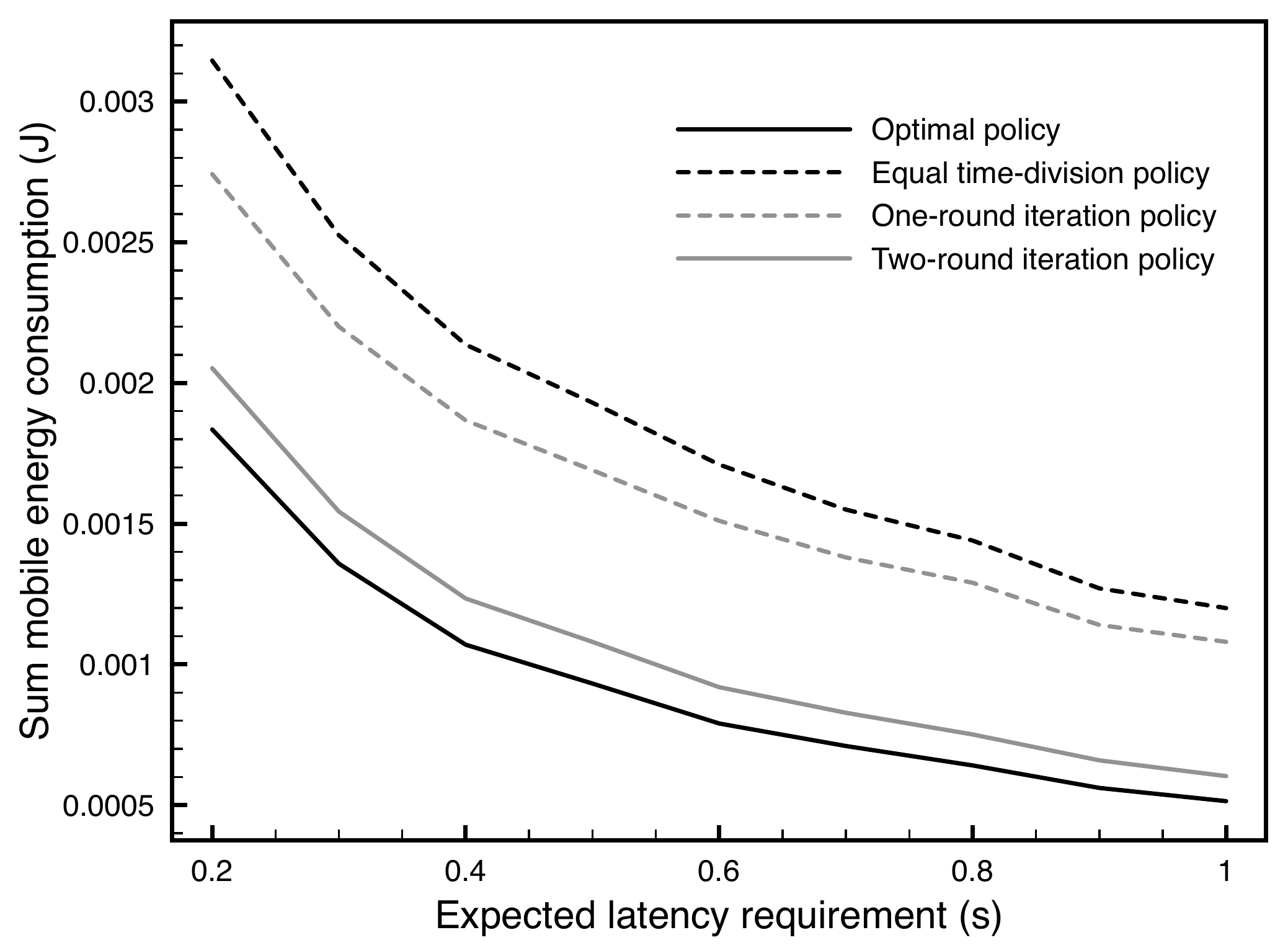}}
\caption{\color{black}{The effects of parameters on the total mobile-energy consumption for the asynchronous MECO resource management  with general arrival-deadline orders.}}
\end{figure} 
First, consider the asynchronous MECO resource management with general arrival-deadline orders. The curves of total mobile-energy consumption versus the monomial order are shown in Fig.~\ref{Fig:GeneEgy_vs_order}. One can observe that the total mobile-energy consumption of the optimal policy grows slowly when the monomial order is small, since computation offloading is preferred to local computing in this regime and the corresponding offloading energy consumption has a small growing rate due to the small $m$. However, after $m$ exceeding a threshold (about $2.5$), the total mobile-energy consumption firstly sees a \emph{fast} and \emph{almost-cubic} increase, and then saturates when the monomial order is large (exceeding about $4$). The reason is that, the energy consumption for offloading and local computing in the moderate-$m$ regime are comparable, both growing at the almost-cubic rates. Nevertheless, in the large-$m$ regime, local computing is more energy-efficient than offloading  and its energy consumption is unaffected by the monomial order. Furthermore, with the energy-efficient time-division control, the optimal policy yields less energy consumption than the equal time-division policy, especially in the moderate-$m$ regime, where about an half energy-consumption reduction is achieved  at  $m\approx3$. {\color{black}{In addition, the one-around and two-round iteration policies have less energy consumption than the equal time-division policy owing to additional data-partitioning and time-division optimizations. The performance of the two-round iteration policy approaches that of optimal policy, indicating that \emph{with more iterations, the iterative algorithm can achieve near-optimal performance.}}} Last, the performance among the four policies converges at a large monomial order due to the gradually ineffectiveness of time-division  control.

Fig.~\ref{Fig:GeneEgy_vs_duration} shows the curves of the total mobile-energy consumption versus the expected required latency. It can be observed that, extending latency requirement can considerably reduce the total mobile-energy consumption for the MECO systems with relatively stringent latency requirements, i.e., small expected required latency; but has less effect when the required latency is already long (exceeding about $0.8$ s). In addition, the optimal policy reduces almost half of energy consumption of the equal time-division policy and thus achieving significant performance gain.

\begin{table*}[t]
  \centering
  \caption{Average Running Time (s) vs. Expected Latency Requirement (s).} \label{Tab:time_vs_latency}
{\color{black}{  \begin{tabular}{ccccccc}
   \hline \hline 
   Expected Latency Requirement   & 0.2 & 0.4 & 0.6 & 0.8 & 1   \\
    \hline\hline
    CVX-based optimal policy   & 5.2939 &  6.4734 & 7.7099  & 8.7704 & 9.7536\\
        \hline
    BCD-based optimal policy   & 0.1005 &  0.1186 & 0.1267 & 0.1309 & 0.1332\\
    \hline
    Equal time-division policy  & 0.0035 &  0.0035 & 0.0035  & 0.0035 & 0.0035\\
    \hline
    One-round iteration policy  & 0.0044 &  0.0045 & 0.0045  & 0.0045 & 0.0045\\
    \hline
    Two-round iteration policy  & 0.0085 &  0.0085 & 0.0085  & 0.0086 & 0.0086\\
    \hline   
     \hline 
  \end{tabular}}}
  \end{table*}

{\color{black}{Last, the computation complexities of different policies are compared by evaluating their average running time using Matlab on a computer equipped with Intel Core i5-4570, 3.20GHz processor and 8GB RAM memory. The results of average running time versus the expected latency requirement are shown in Table~\ref{Tab:time_vs_latency}. It can be observed that the proposed BCD-based policy has much shorter average running time than that using CVX. The reason is that the proposed policy is computed by the iterative algorithm with closed/semi-closed form expression at each iteration, while the latter relies on the universal interior-point method without explicitly exploiting the specific structure of the studied problem. On the other hand, though the BCD-based policy requires longer running time than the baseline policies, it achieves better performance in terms of the total mobile energy consumption (see Fig.~\ref{Fig:GeneEgy_vs_duration}), and the complexity is amenable to  practical implementation.}}

\vspace{-5pt}
\subsection{Identical Arrival-Deadline Orders}
Next, for the asynchronous MECO resource management with identical arrival-deadline orders, the data-arrival time instants and deadlines for the mobiles are generated as follows. First,  a sequence of $(N-1)$ time instants are independently and uniformly generated in the interval of $[0, T]$, where $T$ denotes the total time duration. Next, sorting them in the ascending order and combing it with $(s_0=0)$ and $(s_N=T)$ yields the ordered time instants $\{s_0, \cdots, s_{N}\}$. Then, the data-arrival time instants and deadlines are set as $\l[T_1^{(a)}, T_2^{(a)}, \cdots, T_K^{(a)}\r]=\l[s_0, s_1, \cdots, s_{K-1}\r]$ and $\l[T_1^{(d)}, T_2^{(d)}, \cdots, T_K^{(d)}\r]=\l[s_K, s_{K+1}, \cdots, s_{2K-1}\r]$, respectively.

The impact of expected data size on the total mobile-energy consumption is evaluated in Fig.~\ref{Fig:IdenEgy_vs_data}. The total time duration is set as $T=3$ s. It is observed that, as the expected data size increases, the total mobile-energy consumption of the optimal policy grows at an increasing rate, since both the functions of energy consumption for local computing and offloading are convex and increasing with respect to the data size. Moreover, compared with the baseline policies, the optimal policy has less total mobile-energy consumption and the energy-consumption reduction is more significant for the larger expected data size.

Fig.~\ref{Fig:IdenEgy_vs_duration} depicts the curves of total mobile-energy consumption versus the total time duration. It is observed that extending the total time duration can help reduce the total mobile-energy consumption, since the latency requirements for the mobiles tend to be looser for a larger total time duration. {\color{black}{Moreover, the optimal policy outperforms the baseline policies, especially in the regime with a relatively short total time duration.}}
\begin{figure}[t!]
\centering
\subfigure[Effect of the expected data size.]{\label{Fig:IdenEgy_vs_data}
\includegraphics[width=7.cm]{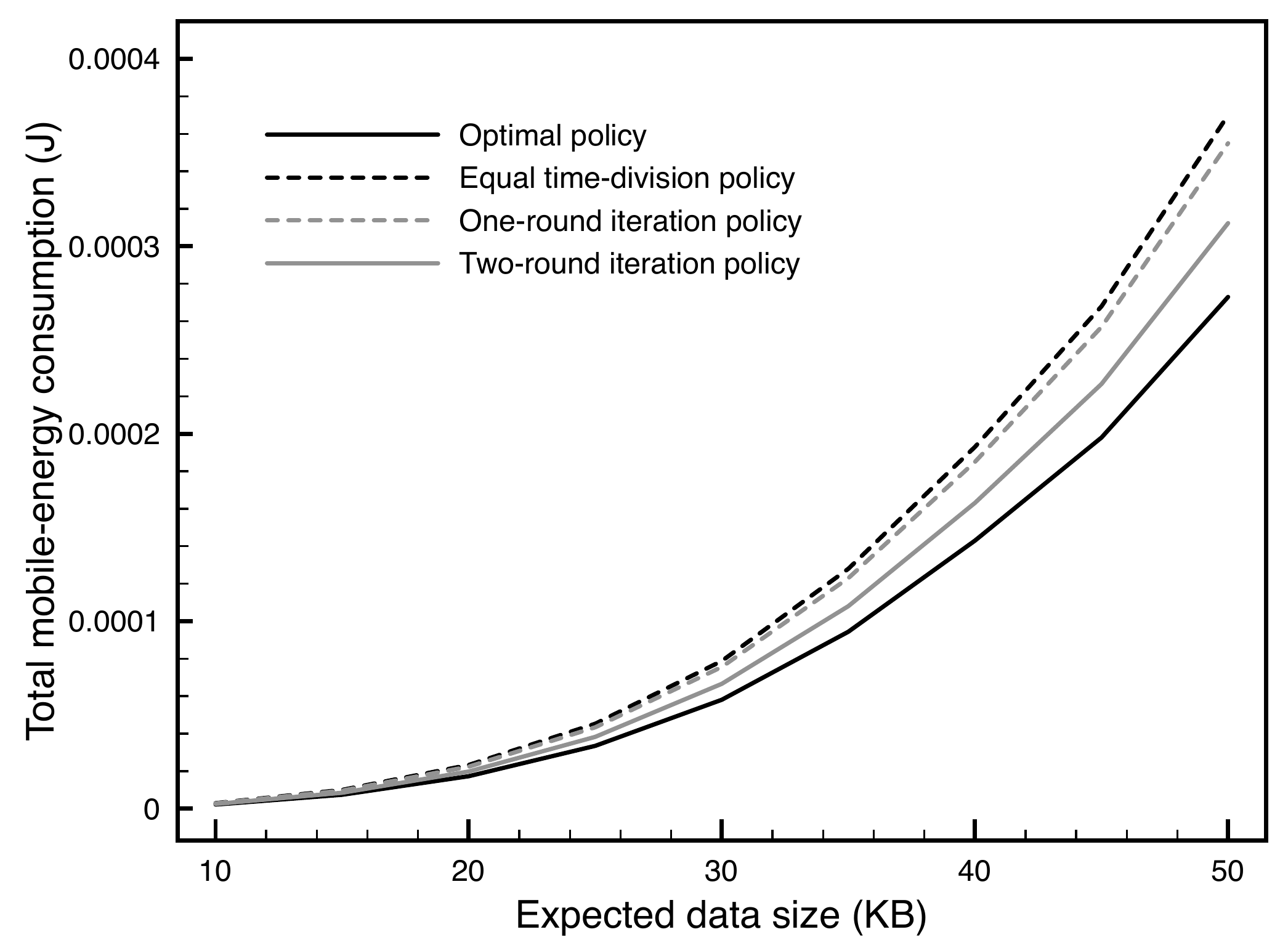}}
\hspace{0.5cm}
\subfigure[Effect of the expected latency requirement.]{\label{Fig:IdenEgy_vs_duration}
\includegraphics[width=7.cm]{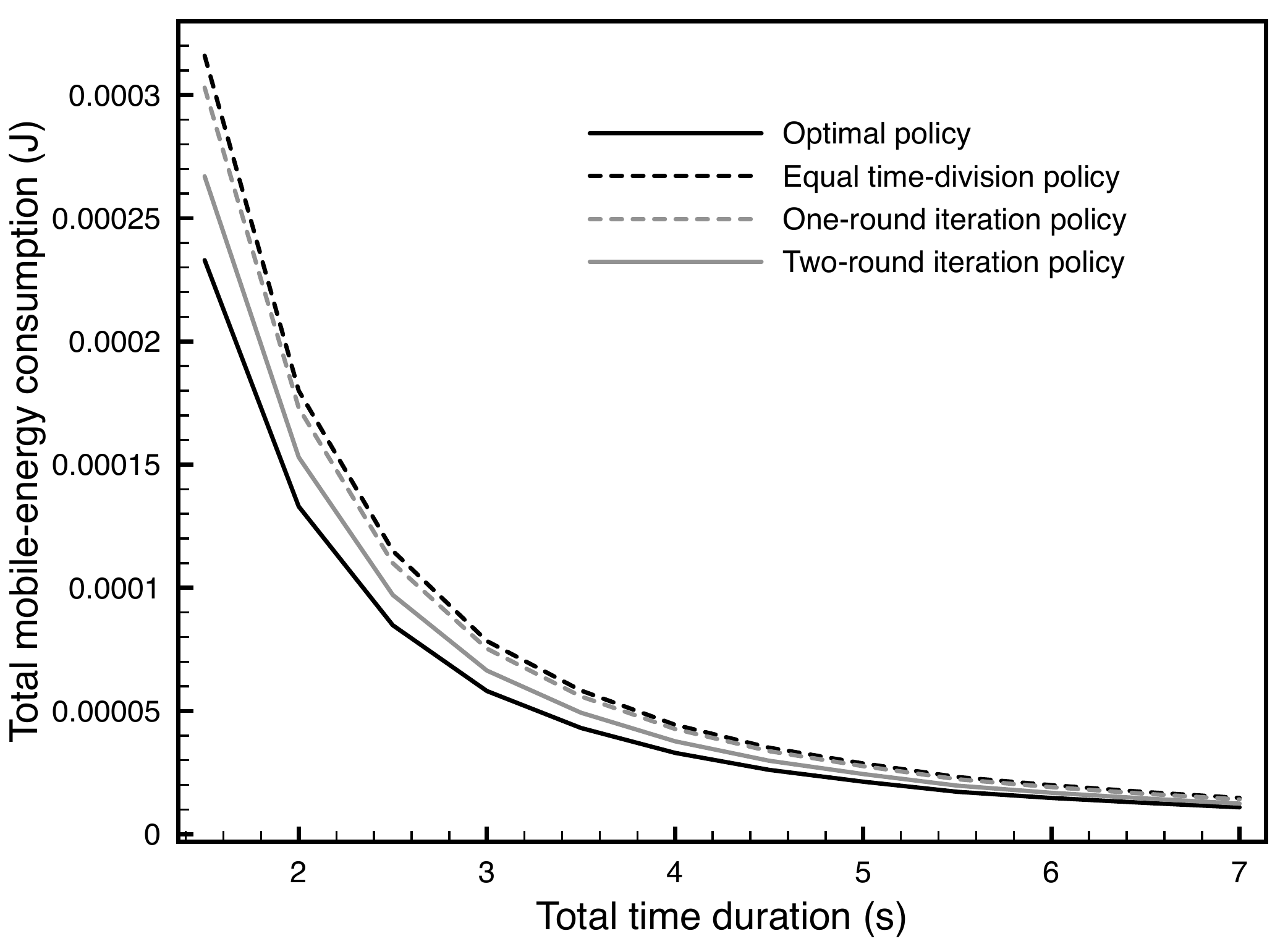}}
\caption{{\color{black}{The effects of parameters on the total mobile-energy consumption for the asynchronous MECO resource management with identical arrival-deadline orders.}}}
\end{figure} 
\vspace{-10pt}
\subsection{Reverse Arrival-Deadline Orders}
Last, we consider the asynchronous MECO resource management with reverse arrival-deadline orders. Similar to the random input-data arrival generation procedure for the case of identical arrival-deadline orders, we first generate a sequence of $(N+1)$ ordered time instants $\{s_0, s_1,\cdots, s_N\}$, where $s_0=0$, $s_N=T$, and $\{s_1,\cdots s_{N-1}\}$ are ordered from the sequence uniformly distributed  over the time interval of $[0,T]$. To form the reverse order, the data-arrival time instants and deadlines are set as $\l[T_1^{(a)}, T_2^{(a)}, \cdots, T_K^{(a)}\r]=\l[s_0, s_1, \cdots, s_{K-1}\r]$ and $\l[T_K^{(d)}, T_{K-1}^{(d)}, \cdots, T_1^{(d)}\r]=\l[s_K, s_{K+1}, \cdots, s_{2K-1}\r]$, respectively.

The curves of total mobile-energy consumption versus the expected data size and total time duration are plotted in Fig.~\ref{Fig:RevEgy_vs_order} and Fig.~\ref{Fig:RevEgy_vs_duration}, respectively. Comparing them with Fig.~\ref{Fig:IdenEgy_vs_data} and Fig.~\ref{Fig:IdenEgy_vs_duration}, we can observe that the total mobile-energy consumption for the case of reverse arrival-deadline orders is much larger than the counterpart with the  identical orders. The reason is that, for the case of the reverse orders, the mobiles that arrive lately have more stringent latency requirements, which contribute to substantial energy consumption. Again, larger performance gain is observed for the larger expected data size and smaller total time duration. Other observations are similar to those from Fig.~\ref{Fig:IdenEgy_vs_data} and Fig.~\ref{Fig:IdenEgy_vs_duration}.
\begin{figure}[t!]
\centering
\subfigure[Effect of the expected data size.]{\label{Fig:RevEgy_vs_order}
\includegraphics[width=7cm]{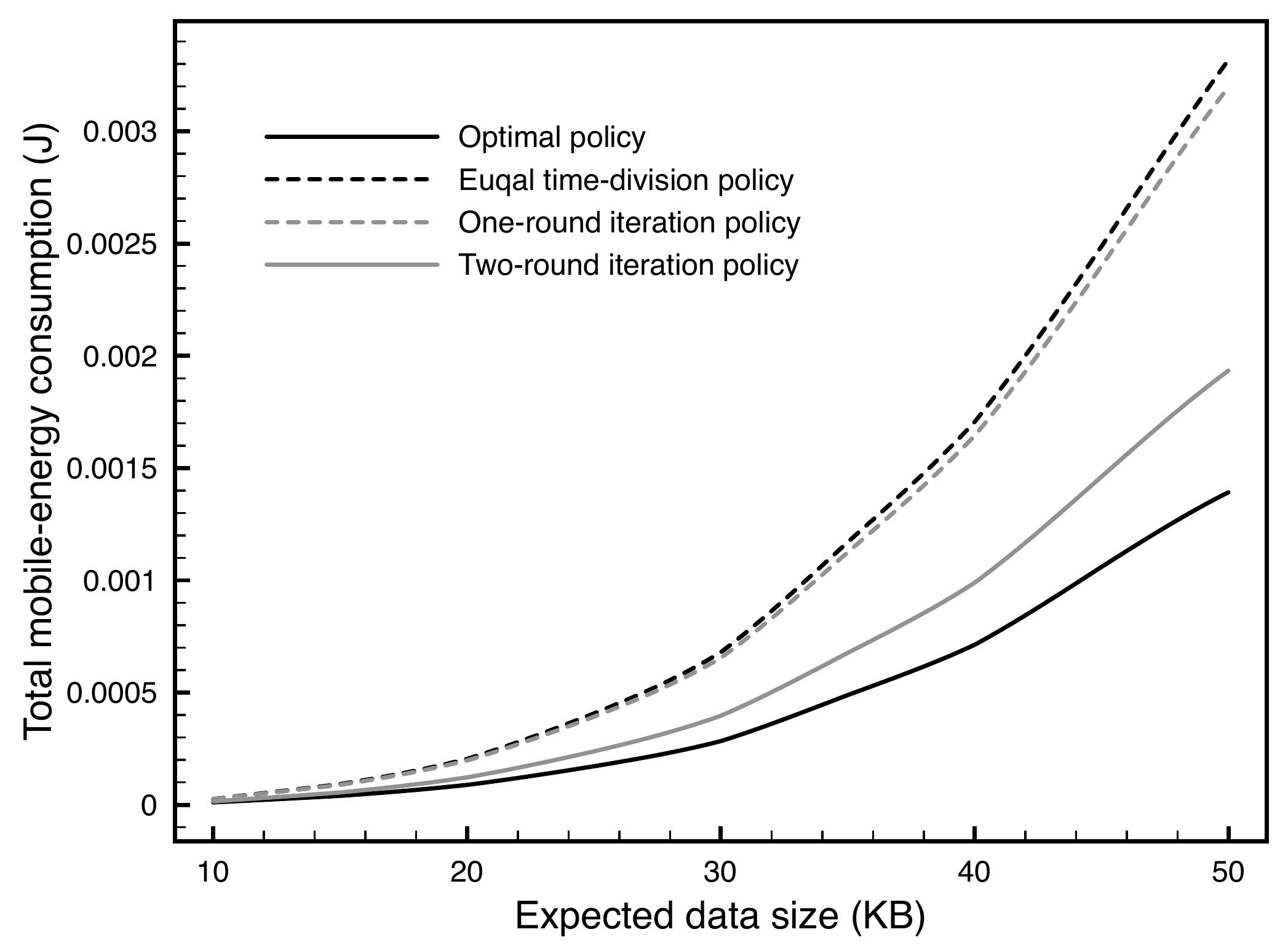}}
\hspace{0.5cm}
\subfigure[Effect of the total time duration.]{\label{Fig:RevEgy_vs_duration}
\includegraphics[width=7cm]{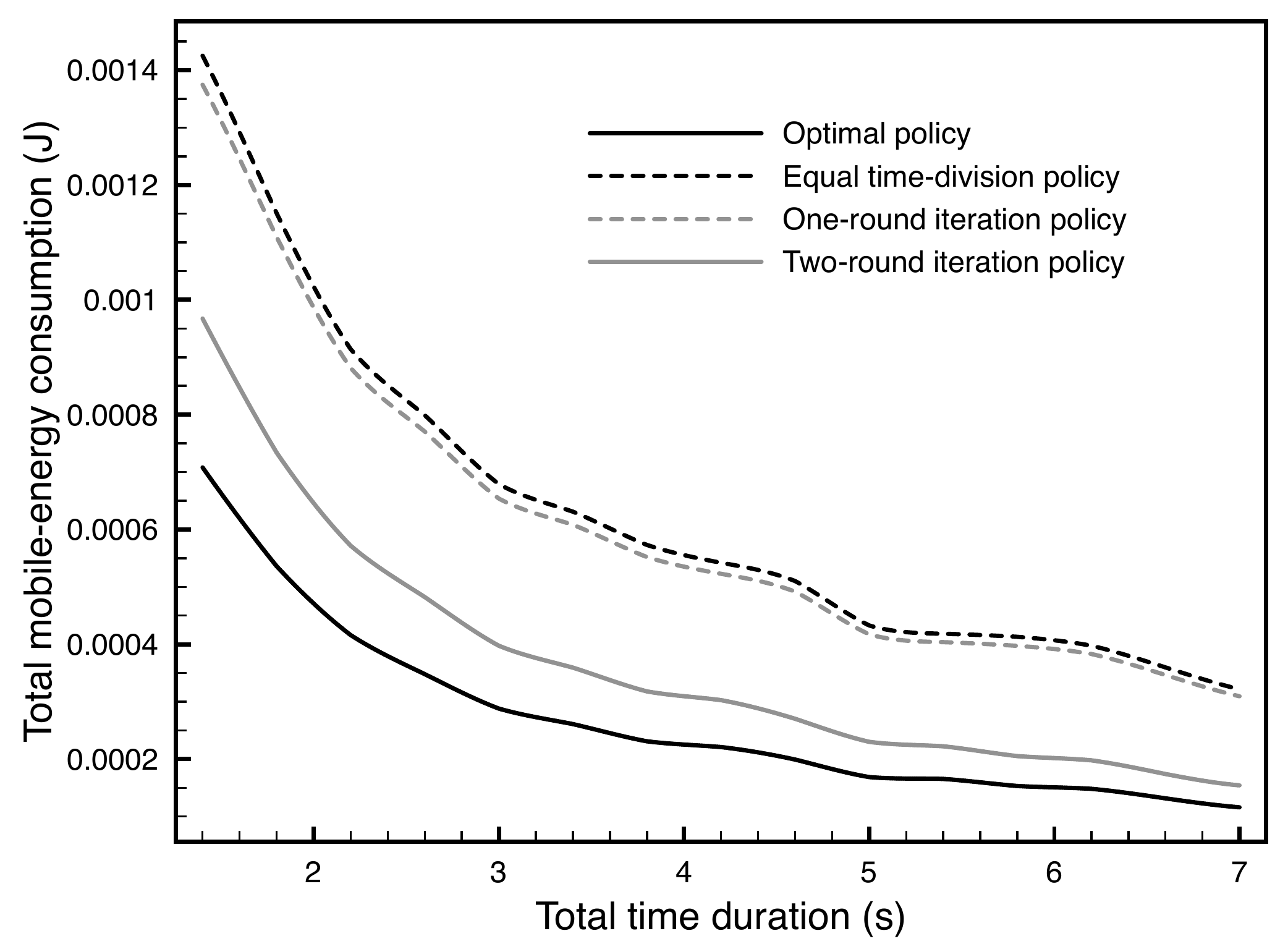}}
\caption{{\color{black}{The effects of parameters on the total mobile-energy consumption for the asynchronous MECO resource management with reverse arrival-deadline orders.}}}
\end{figure} 

\vspace{-10pt}
\section{Concluding Remarks}\label{Sys:Conc}
This paper studies the energy-efficient resource-management  policy for asynchronous MECO systems where the mobiles have heterogeneous data-arrival time instants and deadlines. We first consider the case of general arrival-deadline orders and derive the optimal data-partitioning and time-division policies for minimizing the total mobile-energy consumption by using the BCD method. To obtain more insights into {\color{black}{the structure of the optimal policy}}, we further study the case of identical arrival-deadline orders. To solve the corresponding problem, we first derive the optimal scheduling order and then obtain the optimal joint data-partitioning and time-division policy given the optimal order. Interestingly, it is found that the optimal time-division policy tends to equalize the differences in mobile computation capacities via offloading time allocation to mobiles. The solution approach is extended to another case with reverse arrival-deadline orders. 

\appendix
\vspace{-10pt}
\subsection{Proof of Lemma~\ref{Lem:P1Conv}}\label{App:P1Conv}
Let $\bar{f}(x)$  be defined as $\bar{f}(x)=x^m$. Since $\bar{f}(x)$ is a convex function for $m\geq 1$ over the range of $x\geq 0$, its perspective function $t_k \bar{f}(\ell_k/t_k)=\ell_k^m/t_k^{m-1}$ is also convex for $\ell_k\ge 0$ and $t_k>0$. Using the similar techniques in \cite{wang2008power}, it can be proved that, accounting for both the cases of $t_k>0$ and $t_k=0$, the function of $t_k \bar{f}(\ell_k/t_k)$ is still convex. Thus, the objective function of Problem P1 which is a summation of convex functions, preserves the convexity. Combining it with linear constraints leads to the desired result. 

{\color{black}{\subsection{Proof of Proposition~\ref{Prop:GeneDataParti}}\label{App:GeneDataParti}
First, by relaxing the constraint of Problem P2, we can derive the corresponding solution as $\ell_{k,n}^{'}=h(\xi_k^*)$ for $n\in \bar{A}_k$, where $h(\xi_k^*)$ and $\xi_k^*$ are defined in Proposition~\ref{Prop:GeneDataParti}. Then if $R_k^{\min}\le \sum_{n\in \bar{A}_k} \ell_{k,n}^{'} \le R_k^{\max}$, it means that the solution to Problem P2 is $\ell_{k,n}^{*}=\ell_{k,n}^{'}$. Otherwise, if $\sum_{n\in \bar{A}_k} h(\xi_k^*)<R_k^{\min}$, the solution to Problem P2 should satisfy $\sum_{n\in \bar{A}_k} \ell_{k,n}^{*}=R_k^{\min}$ and thus it can be further derived that $\ell_{k,n}^{*}=\dfrac{t_{k,n}}{\sum_{n\in \bar{A}_k} t_{k,n}} R_k^{(\min)}$. The solution to Problem P2 for the case of $\sum_{n\in \bar{A}_k} h(\xi_k^*)>R_k^{\max}$ can be derived following the similar procedure.}}

{\color{black}{\subsection{Proof of Corollary~\ref{Cor:GeneDataPartiRe}}\label{App:GeneDataPartiRe}
First, note that if $F_k\ge \frac{C_k L_k}{T_k}$ and $D_k\ge L_k C_k$, the constraint of Problem P2 reduces to: $0\le\sum_{n\in \bar{A}_k} \ell_{k,n}\le L_k$. It can be verified that $\ell_{k,n}^*=h(\xi_k^*)$ always satisfies the constraint. 

Next, for $F_k< \frac{C_k L_k}{T_k}$, if $\sum_{n\in \bar{A}_k} h(\xi_k^*)<R_k^{\min}$, then combing it with \eqref{Eq:U_k} leads to $\xi_k^*>(L_k-R_k^{\min})^2$. Since $U_k(\xi_k)$ is monotonically decreasing with $\xi_k$, we have $U(\bar{\xi}_k)<0$ where $\bar{\xi}_k=(L_k-R_k^{\min})^2$, which is equivalent to $\sum_{n\in \bar{A}_k} \ell_{k,n}(\bar{\xi}_k)<R_k^{\min}$. Substituting $\ell_{k,n}(\bar{\xi}_k)$ and $R_k^{\min}$ into it gives 
\begin{equation}\label{Eq:F_k}
\frac{3b_k}{m a_k T_k^2} \l(\frac{T_kF_k}{C_k}\r)^2 \l(\sum\nolimits_{n\in \bar{A}_k} t_{k,n}\r)^{m-1}<\l(L_k-\frac{T_kF_k}{C_k}\r)^{m-1},
\end{equation} which is equivalent to $F_k<\frac{C_k \phi_k}{T_k}$ with $\phi_k$ defined in \eqref{Eq:Phi}. Since $0<\phi_k<L_k$, it can be concluded that the condition of $\sum_{n\in \bar{A}_k} h(\xi_k^*)<R_k^{\min}$ is equivalent to $F_k<\frac{C_k \phi_k}{T_k}$. Following the similar procedure, we can derive that for $D_k< L_k C_k$, the condition of $\sum_{n\in \bar{A}_k} h(\xi_k^*)>R_k^{\max}$ is equivalent to $D_k<C_k (L_k-\phi_k)$. Last, it can be verified that $F_k<\frac{C_k \phi_k}{T_k}$ and $D_k<C_k (L_k-\phi_k)$ cannot be satisfied simultaneously under the problem feasibility condition of $(L_k-\frac{T_kF_k}{C_k}\le \frac{D_k}{C_k})$. Combing the above discussions yields the desired result.
}}

\vspace{-7pt}
\subsection{Proof of Lemma~\ref{Lem:OptOrder}}\label{App:OptOrder}
\vspace{-5pt}
To prove Lemma~\ref{Lem:OptOrder}, we only need to show that, for any optimal scheduling order for Problem P1, it can be transformed to another order in the form of $\{1,1,\cdots, 1,2,2,\cdots, 2,\cdots, K,K,\cdots, K\}$, which is equivalent to $\{1,2,\cdots, K\}$. This argument is proved by construction as follows. Let $\{\ell_{k,n}^*, t_{k,n}^*\}$ denote the optimal solution to Problem P1. Assume $\boldsymbol{\theta}^*=\{\theta_1,\cdots,\theta_{j},\theta_{j+1}, \cdots, \theta_I\}$ is one optimal scheduling order with $\theta_{j}>\theta_{j+1}$. Consider the sub-order of $\{\theta_{j}, \theta_{j+1}\}$. The policy that schedules the subsequent mobile $\theta_j$, followed by $\theta_{j+1}$, satisfies the data causality and deadline constraints. Therefore, we can construct another scheduling sub-order $\{\theta_{j}, \theta_{j+1}\}$, which does not violate the data causality and deadline constraints for both mobiles. In other words, we can construct an alternative scheduling order $\boldsymbol{\theta}^{'}=\{\theta_1,\cdots,\theta_{j+1},\theta_{j}, \cdots, \theta_I\}$. For the newly constructed order, if there exists another sub-order with $\theta_{i}>\theta_{i+1}$, we can switch this sub-order and construct a new order. Repeating this process leads to the desired result.
%

\vspace{-7pt}
\subsection{Proof of Lemma~\ref{Lem:OptOrderRev}}\label{App:OptOrderRev}
This lemma is proved by deriving the optimal scheduling sub-order in the intervals of $\l[0, T_K^{(d)}\r]$ and $\l[T_K^{(d)}, T_K^{(1)}\r]$, respectively. First, consider the time interval $\l[0, T_K^{(d)}\r]$. Assume that each mobile needs to process $\ell_k^*$-bit input data. The scheduling in this duration can be regarded as the case of identical arrival-deadline orders given the same deadline $T_K^{(d)}$. According to Lemma~\ref{Lem:OptOrder}, one optimal scheduling sub-order in this duration is $\{1, 2, \cdots, K\}$. Next, consider the time interval  $\l[T_K^{(d)}, T_K^{(1)}\r]$. For mobiles $1, 2, \cdots, K-1$, it can be regarded that each of them has $(L_k-\ell_k^*)$-bit input data at the same arrival time instant $T_K^{(d)}$, and needs to finish the computation before its individual deadline $T_k^{(d)}$. Using the similar construction approach as presented in Appendix~\ref{App:OptOrder}, we can easily prove that an optimal scheduling sub-order in this duration is $\{K-1, \cdots, 2,1\}$. Combing the two sub-orders together yields the desired result.
%
%
%
%

\end{document}